# A Rechargeable Chromium Battery

Apurva Anjan[1#], Adwitiya Rao[2#], Rohit M. Manoj[1], Vrindha Pongalat[1], Varad Mahajani[3], Sohail Shah[4], Mukesh Bacchav[4], Chandra Veer Singh[2*], and Nikhil Koratkar[1,3*]

[1]Department of Mechanical, Aerospace and Nuclear Engineering, Rensselaer Polytechnic Institute, Troy, NY, 12180, United States

[2]Department of Materials Science and Engineering, University of Toronto, Ontario, ON, M5S-3E4, Canada

[3]Department of Material Science and Engineering, Rensselaer Polytechnic Institute, Troy, NY, 12180, United States

[4]Nuclear Materials Performance Department, Idaho National Laboratory, Idaho Falls, ID 83415, United States

[#]Equal contribution

*Correspondence to: C.V.S. (chandraveer.singh@utoronto.ca) and N.K. (koratn@rpi.edu)

Multivalent ions exchange multiple electrons during redox reactions, leading to the possibility of improved performance. A variety of multivalent ions including Zinc ($Zn^{2+}$), Magnesium ($Mg^{2+}$), Calcium ($Ca^{2+}$), Aluminum ($Al^{3+}$) and Indium ($In^{3+}$) have been deployed in rechargeable batteries with varying degrees of success.[1-9] While Chromium ($Cr^{3+}$) offers superior volumetric capacity (~11,117 mAh $cm^{-3}$) compared to the aforementioned cations, there is no report of a rechargeable Cr battery. This is because Cr metal spontaneously oxidizes forming a passivating oxide[10] that blocks $Cr^{3+}$ ingress and egress. Here we show that this fundamental limitation can be overcome by developing a Cr-rich high-entropy alloy. The alloy consists of five elements (Cr, Bismuth (Bi), Copper (Cu), Tin (Sn), Nickel (Ni)), producing a multi-element native oxide rich in heterointerfaces. Some of these interfaces (such as $Cr_2O_3/Bi_2O_3$) exhibit a very low barrier for $Cr^{3+}$ diffusion, offering a multitude of pathways for efficient $Cr^{3+}$ infiltration and exfiltration, while others (such as $Cr_2O_3/CuO$) block oxygen passage suppressing further oxidation. In a symmetric cell, the Cr alloy supports ~10,000 hours (~5,000 cycles) of reversible Cr insertion/extraction at only ~20 mV overpotential. The Cr-rich alloy anode was also successfully paired with a sulfur cathode to cycle reversibly in a full-cell configuration. These findings could stimulate fundamental study on Cr-ion batteries and high-entropy alloy electrodes, opening-up new pathways for multivalent energy storage.

**Introduction**

Multivalent cations can in principle store more charge than monovalent ions, offering very high theoretical energy densities. For example, Al metal anodes[11] deliver a theoretical capacity of ~8,046 mAh cm$^{-3}$. However, Al$^{3+}$ salts are insoluble in organic electrolytes, forcing use of ionic liquids or aqueous systems that are corrosive, expensive, or limited by oxide formation, hydrogen evolution and narrow voltage windows[12,13]. Cr metal is an attractive alternative multivalent anode (Supplementary Table 1, Supplementary Fig. 1): Cr$^{3+}$ offers a higher theoretical capacity (~11,117 mAh cm$^{-3}$) (Supplementary text) than Al$^{3+}$ and, due to its lower charge density (~505 C mm$^{-3}$ vs ~771 C mm$^{-3}$ for Al) (Supplementary text), could potentially diffuse more readily than Al$^{3+}$. Cr salts are also soluble in certain organic electrolytes, eliminating the need for ionic liquid or aqueous electrolytes, thus circumventing problems like hydrogen evolution. Despite this promise, a Cr based multivalent-ion battery has not been realized to date. The main reason for this is that Cr metal oxidizes forming an ion-blocking $Cr_2O_3$ interface, raising the plating/stripping overpotential and drastically reducing cell voltage and energy density.

To address this challenge, we have developed a Cr-rich high-entropy alloy (Cr-HEA) anode (~40% Cr; ~15% each of Bi, Cu, Ni and Sn). This alloy exhibits a mixed-metal oxide interface, which is highly permeable to passage of Cr$^{3+}$ ions, while effectively blocking oxygen penetration into the sub-surface. *Ab initio* calculations and experiments indicate that a predominantly $Cr_2O_3$ passivation layer (with $Cr_2O_3/Cr_2O_3$ grain boundaries) forms on a conventional Cr metal particle. The diffusion of Cr$^{3+}$ cations is energetically unfavorable across these $Cr_2O_3/Cr_2O_3$ interfaces with a calculated energy barrier ($E_B$) of ~3.83 eV, rendering such interfaces largely impermeable to Cr$^{3+}$ transport. To make matters worse, these $Cr_2O_3/Cr_2O_3$ interfaces are relatively permeable to oxygen passage ($E_B$ of ~0.68 eV), further promoting oxidation and blocking Cr$^{3+}$ migration. By contrast, for Cr-HEA a multitude of elements (namely, Cr, Bi, Cu, Ni and Sn) are present on the surface leading to an abundance of oxide heterointerfaces: such as $Cr_2O_3/Bi_2O_3$, $Cr_2O_3/SnO_2$, $Cr_2O_3/CuO$, and $Cr_2O_3/Cr_2O_3$ among others. *Ab initio* calculations indicate that some of these interfaces such as $Cr_2O_3/Bi_2O_3$ exhibit a very small barrier to Cr$^{3+}$ diffusion (~0.5 eV), while other interfaces such as $Cr_2O_3/CuO$ (~3.65 eV) and $Cr_2O_3/SnO$ (~1.53 eV) retard oxygen permeation. The cumulative effect of this is suppressed oxidation of the Cr-HEA surface, while concurrently allowing facile infiltration/exfiltration of Cr$^{3+}$ cations. Exploiting this, we demonstrate ~10,000 hours (~5000 cycles) of continuous Cr$^{3+}$ insertion/extraction in a symmetric cell with ~20 mV overpotential, indicating that the Cr-HEA oxide interface remains permeable to Cr$^{3+}$ ions and suppresses parasitic reactions. Utilizing this alloy, we demonstrate a rechargeable Cr-HEA||sulfur full-cell battery with reversible cycling performance.

**Results and Discussion**

High-entropy alloys (HEAs) are multicomponent solids (typically ≥5 elements) stabilized by high configurational entropy[14–16]. The Cr-HEA in this work (atomic composition 40:15:15:15:15 of Cr:Bi:Cu:Ni:Sn) was synthesized (see Methods) via ball milling[17] (a scalable powder metallurgy route) (**Fig. 1a**). The estimated configurational entropy for the alloy is 1.504R (where R is the gas constant). The theoretical capacity of Cr-HEA at ~40% Cr loading reduces to ~4,447 mAh cm$^{-3}$.

Further increase in Cr content increases capacity but would lead to excessive distortion due to reduction in configurational entropy – consequently, we limited Cr content in the Cr-HEA to ~40%. Scanning electron microscopy (SEM) shows that the as-milled Cr-HEA powder has an average particle size of ~1.9 ± 0.5 μm (**Fig. 1b,** Supplementary Figs. 2 and 3). X-ray diffraction (XRD) (**Fig. 1c**) shows that before milling, distinct elemental peaks are present, whereas after ~20 hours of milling most elemental peaks merge into a single dominant peak, indicating solid-solution formation[15]. Energy-dispersive X-ray spectroscopy (EDS) (**Fig. 1d**) confirms that Cr, Bi, Cu, Ni and Sn are distributed across these particles, with a Cr content of ~35% (Supplementary Fig. 4). Scanning transmission electron microscopy electron energy loss spectroscopy (STEM-EELS) (Supplementary Fig. 5) similarly detects the presence of all elements other than Bi, which was not observed due to its poor EELS signature. SEM and STEM-EELS of the pure Cr metal powder are provided in Supplementary Figs. 6-7, indicating a particle size range of 20-30 μm. The reduction in size of Cr-HEA relative to Cr particles is attributed to the ball milling synthesis that was used[18].

The chemical composition of the as-synthesized Cr-HEA was probed at atomic resolution using atom probe tomography (APT) (**Fig. 1e**). APT enabled three-dimensional mapping of elemental distributions within a representative nanoscale volume. The analysis verified that Cr, Bi, Cu, Ni, and Sn are all incorporated into the solid solution. However, APT data also reveals that the alloy does not exhibit ideal atomic-scale homogeneity, as highlighted in the overlay image in Fig. 1e. Such local heterogeneities are to be expected for materials produced via high-energy ball milling. X-ray photoelectron spectroscopy (XPS) (**Fig. 1f**) shows that Cr-HEA particles possess a native multi-element oxide on its surface. While the Cr $2p_{3/2}$ spectrum of pure Cr metal particles shows a dominant $Cr_2O_3$ signature (Supplementary Fig. 8), in case of Cr-HEA, the oxide interface contains multiple stable oxides – namely, $Cr_2O_3$, CuO, $SnO_2$, NiO and $Bi_2O_3$. Note that in the Cr $2p_{3/2}$ spectrum, a weak satellite peak (at ~ 578 eV) corresponding to $Cr(OH)_3$ is also observed, but there is no indication of toxic $Cr^{+6}$ being present, which is important from a safety perspective.

Density functional theory (DFT) and nudged elastic band (NEB) calculations via a machine learned interatomic potential (MLIP) were used to understand how $Cr^{3+}$ infiltrates and exfiltrates through Cr-HEA's multielement oxide surface structure. Detailed considerations and methodology for theoretical calculations are available in the methods and supporting information. To evaluate resistance to oxidation of the alloy, binding energy (B.E.)[18] of oxygen atom, on the (110) surface of pure Cr and Cr-HEA (Supplementary Fig. 9) were compared. We isolated 5 distinct sites on the alloy surface to capture the influence of each element on oxidation energetics. The B.E. results are tabulated in Supplementary Table 2. We observed that relative to the B.E. = -2.63 eV (negative energy signifies binding) of pure Cr (110), all sites exhibited a lower binding energy. Site 1, which comprised solely of local Cr surface atoms, showed the highest B.E. = -2.27 eV. Sites 2, 3, 4, 5 exhibited binding energies ranging from -1.26 eV for Cr-Sn competing site (site 3), to -1.85 eV for Cr-Ni competing site (site 5). While these calculations indicate that Cr-HEA exhibits enhanced resistance to oxidation relative to Cr, it is evident that both the pure Cr and Cr-HEA surfaces will oxidize spontaneously when exposed to oxygen, which is consistent with our experimental observations (Fig. 1f, Supplementary Fig. 8).

To study migration of Cr and O species through the native oxide layer on Cr and Cr-HEA, NEB simulations were conducted using the MLIP MACE-MP-0[19], which has shown to exhibit the most accurate determination of activation barriers, relative to other universal MLIPs, as shown by Deng et al.[20]. Due to presence of multiple elements on the Cr-HEA surface, the oxide surface on Cr-HEA would likely comprise of the most stable oxides, phase separated from each other. Further, due to Cr content in the alloy being the highest, most of the phase separated oxide interfaces would comprise of a $Cr_2O_3$ surface, with another elemental oxide. To evaluate which oxide interfaces are stable, interface reactivity calculations were conducted to obtain interface reaction energies for different oxide combinations with $Cr_2O_3$ via the interface reaction module of *Materials Project*[21] and the results are tabulated in Supplementary Table 3. The interface reaction energies showed that $Bi_2O_3$, $SnO_2$ and $CuO$ develop stable interfaces with $Cr_2O_3$. By contrast, $Cu_2O$ and $NiO$, exhibited a nonzero interface reaction energy, indicating that these oxides would react with $Cr_2O_3$ to form a mixed oxide. For the scope of this analysis, we have avoided mixed oxide interfaces with $Cr_2O_3$ for two reasons: 1) XPS (**Fig. 1f**) indicated that CuO is the dominant oxide, so the number of $Cr_2O_3/Cu_2O$ interfaces if any would be insignificant; 2) NiO reacts with $Cr_2O_3$ to yield $Cr_2NiO_4$ (Supplementary Table 3). Since $Cr_2O_3$ gets consumed to yield this composition, occurrence of $Cr_2O_3/Cr_2NiO_4$ interfaces would be less likely as compared to the other stable oxide configurations discussed above. Hence, we focus our studies on $Cr_2O_3/Bi_2O_3$, $Cr_2O_3/SnO_2$ and $Cr_2O_3/CuO$ hetero-interfaces as well as conventional $Cr_2O_3(001)/Cr_2O_3(001)$ grain boundary interfaces.

NEB results for the computed energy barrier ($E_B$) for both Cr and O migration across the various interfaces, are presented in **Figs. 2a**, and **2c**. Our results for the $Cr_2O_3(001)/Cr_2O_3(001)$ grain boundary interface match the experimental findings of Sabioni et al[22], with O showing a much higher propensity for ease of migration ($E_B$ = 0.68 eV) when compared to Cr ($E_B$ = 3.83 eV). This high barrier for Cr migration explains why a large overpotential would be required to drive Cr transport through the oxidized Cr metal surface. For the Cr-HEA surface (**Fig. 2b**), all three stable hetero-interfaces, i.e. $Cr_2O_3/Bi_2O_3$ ($E_B$ = 0.5 eV), $Cr_2O_3/SnO_2$ ($E_B$ = 2.04 eV) and $Cr_2O_3/CuO$ ($E_B$ = 2.05 eV) show a decidedly reduced migration energy barrier relative to $Cr_2O_3/Cr_2O_3$ ($E_B$ = 3.83 eV). In particular, $Cr_2O_3/Bi_2O_3$ interfaces offer highly efficient channels for Cr diffusion, leading to minimal electrochemical energy loss during cycling. The picture is completely reversed, when one considers O diffusion (**Fig. 2d**). All three stable hetero-interfaces, i.e., $Cr_2O_3/Bi_2O_3$ ($E_B$ = 0.95 eV), $Cr_2O_3/SnO_2$ ($E_B$ = 1.53 eV) and $Cr_2O_3/CuO$ ($E_B$ = 3.65 eV) show a markedly larger migration energy barrier relative to $Cr_2O_3/Cr_2O_3$ ($E_B$ = 0.68 eV). Thus in Cr-HEA, oxidation of the underlying alloy is suppressed, while $Cr^{3+}$ can be efficiently inserted and extracted through the multi-element native oxide layer.

To compare $Cr^{3+}$ insertion/extraction from Cr-HEA vs Cr metal, we fabricated electrodes by mixing Cr-HEA or Cr powders with PVDF binder and Super P carbon (8:1:1 by weight) in N-methyl-2-pyrrolidone and cast the slurry onto carbon-fibre current collectors. The resulting Cr-HEA electrode exhibits a porous microstructure (Supplementary Fig. 10), and EDS mapping confirms (Supplementary Fig. 11) presence of all five constituent elements (Cr, Bi, Cu, Ni, Sn) in the electrode. We assembled symmetric coin cells in an Ar atmosphere using ~2.0 M chromium trichloride hexahydrate powder in dimethyl sulfoxide (DMSO) solvent as the electrolyte. Chromium trichloride hexahydrate was chosen as the salt for its low cost and good solubility (Supplementary Table 4), while DMSO was selected for its high dielectric constant, strong donor number, and high boiling point (Supplementary Table 5), which stabilizes solvated $Cr^{3+}$ cations[23]. Note that the Cr salt used in this study is utilized at industrial scale as a catalyst, mordant in textiles, for water treatment and in chemical synthesis. At reduced salt concentrations of ~0.5 and ~1.0 M, we observed electrolyte degradation (Supplementary Fig. 12-14); which was not observed at a higher concentration of ~2.0 M and above. Besides this, the ~2.0 M electrolyte was also selected for all the further studies due to its enhanced voltage stability window (Supplementary Fig. 15).

Symmetric cells with Cr metal electrodes did not cycle well as indicated in **Fig. 3a** and **3b**. Because of the ion-blocking $Cr_2O_3$ layer on Cr metal, a large overpotential is needed to force the insertion and extraction of $Cr^{3+}$ ions through this interface. For example, at a current density of ~0.2 mA cm$^{-2}$, over 1.5 V is needed to force $Cr^{3+}$ through the oxide layer on the Cr metal surface. By contrast, for Cr-HEA the potential needed to drive $Cr^{3+}$ through the multi-element oxide layer is about 20 mV. This voltage remains stable over 5,000 cycles, which corresponds to > 1 year (~10,000 hours) of continuously shuttling Cr ions back and forth. The superiority of Cr-HEA is retained over a range of current densities (**Fig. 3c**, Supplementary Fig. 16). Moreover, an asymmetric voltage profile is observed for the Cr ∥ Cr cell (Supplementary Fig. 17), which is indicative of uneven plating/stripping. Conversely, the symmetric voltage response recorded for the Cr-HEA ∥ Cr-HEA cell indicates uniform $Cr^{3+}$ shuttling. Uneven plating/stripping is also reflected in non-uniform columbic efficiency (Supplementary Figs. 18 and 19) of Cr in comparison to the Cr-HEA symmetric cell. We also explored electrolytes with higher salt concentrations of ~3.0 M and ~4.0 M $CrCl_3.6H_2O$ in DMSO (Supplementary Fig. 20), however the ~2.0 M salt gave the best performance. The Cr-HEA symmetric cell was also cycled at an elevated areal capacity of ~6.0 mAh cm$^{-2}$ (Supplementary Fig. S21), achieving far smaller voltage hysteresis (overpotential) compared to the Cr cell, despite the fact that the baseline Cr cell was operated at ~30-fold lower areal capacity (~0.2 mAh cm$^{-2}$).

To quantify how much Cr can be removed from Cr-HEA without encountering structural collapse, we have performed a one-way Cr-extraction test (Methods) and concluded that ~37% Cr can be extricated while keeping the HEA structure relatively intact. When we attempted to extract ~50% of the total Cr present in Cr-HEA, the XRD pattern showed a major change in peaks, presumably caused by a phase change in Cr-HEA (**Fig. 3d**). For the areal capacities utilized in Fig. 3, the amount of Cr extracted is < 37%, hence we do not expect the structure to undergo major modifications. This was verified by conducting *ex situ* XRD measurements of the cycled Cr-HEA electrode after 0, 50, 100, 250 and 1000 cycles. The results confirm that the XRD patterns of the Cr-HEA electrode during the cycling process do not change significantly indicating that the alloy structure remains relatively intact even after long hours of continuous $Cr^{3+}$ shuttling (**Fig. 3e**).

To demonstrate that the Cr-HEA anode could be deployed in a full-cell device, we paired it with a sulfur (S) cathode (**Fig. 4a**). This configuration represents the first practical demonstration of a rechargeable Cr battery chemistry. Since S exhibits poor electronic conductivity, a S/carbon composite cathode was utilized (Methods). The electrochemical performance was first evaluated using cyclic voltammetry (CV) in a voltage window of 0.05–1.0 V at ~0.5 mV s$^{-1}$ (**Fig. 4b**). The CV profile exhibits two distinct cathodic peaks at ~0.6 V (A) and ~0.1 V (B), corresponding to the two-step reduction of elemental sulfur (S$_8$) to long-chain chromium polysulfides (Cr$_x$S$_n$ 4≤n≤8) and subsequent short chain polysulfides (Cr$_x$S$_n$ 3≤n≤1). In the anodic scan, two corresponding oxidation peaks at ~0.4 V (C) and ~0.8 V (D) indicate the reversible conversion back to long-chain polysulfides and finally to S$_8$, consistent with the established mechanism for S-based cathodes[24]. Galvanostatic cycling at a current density of ~100 mA g$^{-1}$ was conducted (**Fig. 4c-d**). The cell delivered an initial specific capacity of ~300 mAh g$^{-1}$, retaining ~255 mAh g$^{-1}$ after 150 cycles, which corresponds to a capacity retention of ~83%. The coulombic efficiency averaged over 150 cycles is ~94.5%. The observed capacity fade is likely related to polysulfide dissolution and shuttling, a common problem with S-based cathodes.

To elucidate the Cr$^{3+}$ storage mechanism within the S cathode, we performed *ex situ* XPS on the cathode prior to cycling and in the fully discharged state. The S 2p spectrum of a fully discharged cathode (0.05 V) reveals the emergence of a new doublet at ~161.2 and ~162.3 eV, which is shifted (**Fig. 4e**) with respect to the pristine S electrode. This feature is characteristic of metal-sulfur bond formation, confirming conversion of S$_8$ to chromium sulfides. This conclusion is strongly corroborated by the high-resolution Cr 2p spectrum (**Fig. 4f**). The pristine S cathode is devoid of any Cr signature. In the discharged state, the S cathode exhibits a prominent Cr$^{3+}$ peak (~576.5 eV) (corresponding to Cr$_2$S$_3$) providing direct evidence of Cr$^{3+}$ migration from the Cr-HEA anode to the S$_8$ cathode to participate in the conversion reaction. This Cr migration was further corroborated by *ex situ* EDS mapping, which indicates that the pristine cathode (**Fig. 4g**) is devoid of Cr, while the fully discharged cathode (**Fig. 4h**) displays a uniform distribution of both S and migrated Cr elements across the electrode surface.

Critically, a control full-cell assembled with a pure Cr-metal anode and the same S-cathode showed negligible electrochemical activity. This confirms that while pure Cr has a higher theoretical capacity, its ion-blocking native oxide renders it unusable. In contrast, Cr-HEA, with its ion-conductive and stable multi-metal surface oxide layer, successfully enables reversible Cr-ion transport, unlocking the first practical chromium rechargeable battery. Looking ahead, future advancement of this technology will hinge on increasing the Cr content in Cr-HEA electrodes to boost capacity as well as the discovery of improved high-voltage cathodes and electrolytes that can be paired with low-voltage Cr-HEA anodes.

**Methods:**

*Materials:* Cr (99% trace metal basis, powder < 45 µm), Bi (99.99% trace metal basis), Cu (99.99% trace metal basis), Sn (99.5% trace metal basis, < 150 µm), Ni (99.7% trace metal basis, < 50 µm) elemental powders, Chromium (III) chloride, DMSO, GF/D and GF/C separator, toluene, were purchased from Sigma Aldrich. Conductive carbon paper was obtained from MSE Supplies. Graphite powder (99.99% trace metal basis, < 45 µm) and Super P carbon was obtained from Sigma Aldrich. Sulfur/carbon composite powder was procured from MTI supplies.

*Synthesis of Cr-HEA powder:* The elemental powders were mixed in an approximate atomic ratio of 40:15:15: 15:15. The powders were mechanically alloyed in a high energy ball milling machine (Retsch PM-100). Ball to powder ratio of ~10:1 was maintained, and toluene was used as the process control agent. The milling was carried out for ~40 hours at ~600 RPM. The resulting powder was dried to evaporate toluene and is named as Cr-HEA. As a control sample Cr powder was used as-received.

*Characterizations:* The resulting working electrodes were characterized structurally using field emission scanning electron microscopy (FE-SEM) images captured on a Carl Zeiss SUPRA instrument. STEM experiments were done on a Thermo Fisher Scientific Titan F30 operating at 300kV with a convergence angle of 19mrad. XRD was performed using a PANalytical X'Pert Pro Diffractometer with Cu-Kα radiation (λ = 1.54 Å). XPS measurements were done on PHI Versa Probe II XPS, and the data were analyzed on Multipak software. The vacuum degree of the analysis chamber was < $5 \times 10^{-9}$ torr. The testing samples were transferred to the inlet chamber through an air-isolated chamber. All the binding energies were calibrated with the C1s peak at 284.8 eV. APT specimens were prepared using focused ion beam milling using a standard lift-out procedure on a dual beam FEI SEM. Multiple tips were prepared due to high porosity within the liftout location. The atom probe analysis was done using the Cameca Local Electrode Atom Probe (LEAP) 5000XR instrument. Samples were cooled to a base temperature of 50K and all samples were run in laser mode with laser energies varying from 70-100 pJ. A pulse repetition rate of 125 kHz and a detection rate of 0.005 atoms/pulse was used for all acquisitions. Datasets were reconstructed using the AP Suite6 software from CAMECA inc. Due to varying evaporation fields of multiple phases and the abrupt fluctuations of voltage during acquisitions, reconstructions were carried out using the fixed shank angle method[25]. SEM images taken after final sharpening of needles were used to estimate the shank angle for reconstructions.

*Battery Assembly:* Cr-HEA powder was ground with super P carbon, polyvinylidene difluoride (PVDF) and N-Methyl-2-pyrrolidone (NMP) according to weight ratio of ~8:1:1. The obtained homogenous slurry was casted onto carbon fiber paper (CFP) with a doctor blade. After coating the electrode was dried at ~60 °C for ~12 h and cut into round sheets with an electrode area of ~1.27 $cm^2$. The Cr-HEA loading on CFP was ~17 mg $cm^{-2}$, which corresponded to Cr loading of ~6.8 mg $cm^{-2}$. Pure Cr electrodes were also coated using the same procedure with a mass loading of ~6 mg $cm^{-2}$. 2032-type coin cells were assembled in an Ar-filled glovebox with $H_2O$ < 1ppm and $O_2$ < 1ppm. For symmetric cells, Cr-HEA electrodes were used as both electrodes with

Whatman Glass fiber (GF/C) separator and ~80 µL of ~2M (mol L$^{-1}$) CrCl$_3$.6H$_2$O as electrolyte. For cathode preparation, sulfur/carbon composite powder was dried in a vacuum oven at ~120°C for ~2h to remove excess moisture. The cathode was obtained by grounding the sulfur/carbon powder with Super P carbon, and PVDF (weight ratio of 8:1:1) in NMP. The obtained homogenous slurry was cast onto CFP with a doctor blade. After coating the cathode was dried at ~60 °C for ~12 h. The cathode mass loading was ~1.2 mg cm$^{-2}$. For full cell testing, we used ~80 µL of ~2M (mol L$^{-1}$) CrCl$_3$.6H$_2$O as the electrolyte and a GF/D separator.

***Electrochemical Measurements***: Galvanostatic Cr plating/stripping (discharging ~1 hour, rest ~1 min, and charging ~1 hour) were performed on the Neware CT-4008T battery testing system at various current densities. For the full cell, galvanostatic charge/discharge cycles were performed over the voltage range of 0.05 to 1V and all gravimetric numbers for current density and specific capacities are based on the mass of sulfur. EIS data were collected in the frequency range of ~0.1 Hz to ~0.1 MHz on a Gamry Interface 1010e instrument. Linear Sweep Voltammetry was performed in a three-electrode setup with Titanium foil (working electrode), Ag/AgCl (reference electrode) and Platinum (counter electrode). The one-way Cr extraction test was performed with Cr-HEA and Titanium foil as the electrodes, 2M CrCl$_3$.6H$_2$O in DMSO electrolyte and GF/C separator. The test was done by fixing the areal capacity at ~1 mAh cm$^{-2}$.

***Density Functional Theory Calculations***: All first principles calculations were carried out within the density functional theory framework using the Vienna Ab Initio Simulation Package (VASP). Core electrons were represented with projector augmented wave pseudopotentials, and exchange correlation effects were described using the generalized gradient approximation with the Perdew Burke Ernzerhof (PBE) functional[26]. A plane wave basis with a kinetic energy cutoff of 550 eV was employed. All BCC surfaces were modeled as 72 atom slabs composed of four layers and separated from periodic images by a vacuum region of approximately 15 Å. Brillouin zone sampling used a 3x3x1 Monkhorst Pack grid. Structural relaxations proceeded until the total energy change was below 1x10$^{-5}$ eV and residual forces were below 0.02 eV/ Å. Dispersion interactions were included using the DFT-D3 method[27]. The oxygen binding energy, defined as the energy gained upon forming an oxygen bond on the surface, was computed using the expression E − (E(surface+oxygen) + E(H$_2$O) – E(H$_2$)). All structures were visualized using the VESTA package.

***Nudged Elastic Band calculations (DFT)***: All NEB calculations were performed with the MACE-MP-0 potential[19], which is trained on the MPTrj dataset, comprising of all the target compositions simulated in this study[28]. The python-based NEB module of atomic simulation environment (ASE) was used to conduct the NEB calculations with the FIRE optimizer and a force convergence parameter of 0.02 eV/ Å [29]. Bulk atoms of the interfaces were constrained during optimization and NEB calculations, and 9 images were used to compute the migration barrier, with all the migration paths being under 5 Å.

***Simulation Model***: To determine the oxidation energetics of the HEA alloy anode, a 72 atom BCC (110) surface was created for the composition $Cr_{0.347}Bi_{0.152}Sn_{0.166}Cu_{0.166}Ni_{0.166}$ which is comparable to the experimental configuration (Supplementary Fig. 4). A solid solution phase was obtained by randomly shuffling the atoms to obtain a highly mixed configuration and the lattice parameter of 2.605 Å was obtained through iterative optimization. The optimized structure exhibited some degree of distortion primarily due to the multi-element nature of the composition, high Cr content and a small representative structure for DFT calculations. For all the interface models, the (100) surface was considered since it is the most stable surface[10] for $Cr_2O_3$ and for the majority of other oxides. All interface structures had a < 5% mismatch in lattice parameters with a vacuum of 15 Å to avoid periodic interactions.

## Acknowledgments

N.K. acknowledges funding support from the John A. Clark and Edward T. Crossan Chair Professorship at the Rensselaer Polytechnic Institute. STEM was performed in collaboration with Dr. Matthew Boebinger at the Center for Nanophase Materials Sciences, which is a US Department of Energy, Office of Science User Facility at Oak Ridge National Laboratory. APT measurements were performed at Idaho National Laboratory. The authors would like to thank Dr. Fudong Han, Dr. Anil Pathak, Dr. Vikram Kishore Bharti and Ankit Chaurasiya for insightful discussions. We also acknowledge facilities at the MNCR clean room and CMDIS facilities at the Rensselaer Polytechnic Institute.

## Author contributions

AA and NK envisioned the concept and developed the research plan. NK and CVS supervised the project. AA synthesized and characterized the Cr-HEA and baseline Cr materials. He also conducted all physical and electrochemical characterizations. AR and CVS performed all first principles calculations. RMM, VM and VP contributed to materials characterization and data analysis. SS and MB conducted the APT experiments. AA, AR and NK wrote the manuscript.

*Competing interests:* Authors have no competing interests

# Supplementary Information

## A Rechargeable Chromium Battery

Apurva Anjan[1#], Adwitiya Rao[2#], Rohit M. Manoj[1], Vrindha Pongalat[1], Varad Mahajani[3], Sohail Shah[4], Mukesh Bacchav[4], Chandra Veer Singh[2*] and Nikhil Koratkar[1,3*]

[1]Department of Mechanical, Aerospace and Nuclear Engineering, Rensselaer Polytechnic Institute, Troy, NY, 12180, United States

[2]Department of Materials Science and Engineering, University of Toronto: Ontario, ON, M5S-3E4- Canada

[3]Department of Material Science and Engineering, Rensselaer Polytechnic Institute, Troy, NY, 12180, United States

[4]Nuclear Materials Performance Department, Idaho National Laboratory, Idaho Falls, ID 83415, United States

[#]Equal contribution

*Correspondence to: C.V.S. (chandraveer.singh@utoronto.ca) and N.K. (koratn@rpi.edu)

# Supplementary Text

**Theoretical capacity calculations for Chromium:**

Theoretical gravimetric capacity, $Q = \frac{n \times 26800}{MW}$ mAh g$^{-1}$

n is the number of mol of electrons transferred by 1 mol electrode

MW is the molecular weight of 1 mol of electrode, g/mol

$Q_{Cr} = \frac{3 \times 26800}{51.996} = 1546.27$ mAh g$^{-1}$

Theoretical volumetric capacity (mAh cm$^{-3}$) = gravimetric capacity (mAh/g) × density (g cm$^{-3}$)

Theoretical volumetric capacity (mAh cm$^{-3}$) = 1546.27 mAh g$^{-1}$ × 7.19 g cm$^{-3}$ = 11117.7 mAh cm$^{-3}$

**Ionic charge density calculation for Chromium:**

$Cr^{3+}$ ion's charge: 3 × e, e is the elementary charge ($1.602 \times 10^{-19}$C)

$Cr^{3+}$ ion charge = $4.806 \times 10^{-19}$C

$Cr^{3+}$ ion size: ~61pm ($61 \times 10^{-12}$m), volume = 4/3 × π × r$^3$

Volume = $\frac{4}{3} \times 3.14159 \times (61 \times 10^{-12})^3 \approx 9.5 \times 10^{-31}$ $m^3 \approx 9.5 \times 10^{-22}$ $mm^3$

Charge density ($\rho$) = $\frac{4.806 \times 10^{-19}}{9.5 \times 10^{-22}} = 505.89$ $C\ mm^{-3}$

**Ionic charge density calculation for Aluminum:**

$Al^{3+}$ ion's charge: 3 × e, e is the elementary charge ($1.602 \times 10^{-19}$C)

$Al^{3+}$ ion charge = $4.806 \times 10^{-19}$C

$Al^{3+}$ ion size: ~53 pm ($53 \times 10^{-12}$m), volume = 4/3 × π × r$^3$

Volume = $\frac{4}{3} \times 3.14159 \times (53 \times 10^{-12})^3 \approx 6.23 \times 10^{-31}$ $m^3 \approx 6.23 \times 10^{-22}$ $mm^3$

Charge density ($\rho$) = $\frac{4.806 \times 10^{-19}}{6.23 \times 10^{-22}} = 771.43$ $C\ mm^{-3}$

# Supplementary Tables

**Supplementary Table 1** | Properties of all metals used as anode materials along with Chromium: abundance, density, redox potential, theoretical specific capacity and energy density, ionic radius and charge density.

| Element | Abundance in Earth's crust (%) | Density (g/cm$^3$) | Redox Potential (vs SHE) | Theoretical specific capacity (mAh g$^{-1}$) | Energy density (mAh cm$^{-3}$) | Ionic radius (Å) |
|---|---|---|---|---|---|---|
| Lithium | 0.0017 | 0.53 | -3.04 | 3861 | 2061 | 0.76 |
| Zinc | 0.0078 | 7.14 | -0.76 | 819 | 5847 | 0.74 |
| Chromium | 0.014 | 7.19 | -0.9 | 1546 | 11117 | 0.61 |
| Potassium | 1.5 | 0.856 | -2.93 | 686 | 587 | 1.33 |
| Sodium | 2.3 | 0.97 | -2.71 | 1166 | 1131 | 1.02 |
| Magnesium | 2.9 | 1.74 | -2.35 | 2205 | 3836 | 0.72 |
| Calcium | 5 | 1.54 | -2.87 | 1337 | 2058 | 0.99 |
| Iron | 6.3 | 7.86 | -0.44 | 960 | 7545 | 0.76 |
| Aluminum | 8.1 | 2.70 | -1.66 | 2980 | 8046 | 0.53 |

**Supplementary Table 2 |** Binding energy of oxygen atom on different Cr-HEA sites.

| Site index | Site/system | Binding energy (eV) |
|:---:|:---:|:---:|
| 0 | Pure Cr (110) surface | -2.63 |
| 1 | Cr-Cr | -2.27 |
| 2 | Cr-Bi | -1.38 |
| 3 | Cr-Sn | -1.26 |
| 4 | Cr-Cu | -1.74 |
| 5 | Cr-Ni | -1.85 |

**Supplementary Table 3** | Interface reactivity values.

| Interface | Interface Reaction Energy (eV/atom) | Phase equilibrium compositions |
|---|---|---|
| $Cr_2O_3$-$Bi_2O_3$ | 0 | - |
| $Cr_2O_3$-$SnO_2$ | 0 | - |
| $Cr_2O_3$-$CuO$ | 0 | - |
| $Cr_2O_3$-$Cu_2O$ | -0.064 | $CrCuO_2$ |
| $Cr_2O_3$-$NiO$ | -0.002 | $Cr_2NiO_4$ |

**Supplementary Table 4 |** Price comparison of Chromium salts

| Salt | Price (USD/100gm) | Cas number |
|---|---|---|
| Chromium (III) chloride hexahydrate | 26.40 | 10060-12-5 |
| Chromium (II) chloride | 2460 | 10049-05-5 |
| Chromium (III) chloride | 1928 | 10025-73-7 |
| Chromium (III) nitrate nonahydrate | 72 | 7789-02-8 |
| Chromium (III) sulfate hydrate | 93.70 | 15244-38-9 |
| Chromium (III) fluoride tetrahydrate | 157 | 123333-98-2 |

**Supplementary Table 5 | Properties of commonly reported solvents in batteries.**

| Solvent | Dielectric constant | Donor number | Boiling point (°C) |
|---|---|---|---|
| DMSO | 46.45 | 29.8 | 189 |
| AN | 37.5 | 14.1 | 82 |
| Diethyl ether | 4.33 | 19.2 | 35 |
| DME | 7.2 | 20 | 85 |
| EC | 89.78 | 16.4 | 243 |
| PC | 64.92 | 15.1 | 242 |
| DMC | 3.107 | 16 | 90 |
| Triglyme | 7.5 | 20 | 216 |

# Supplementary Figures

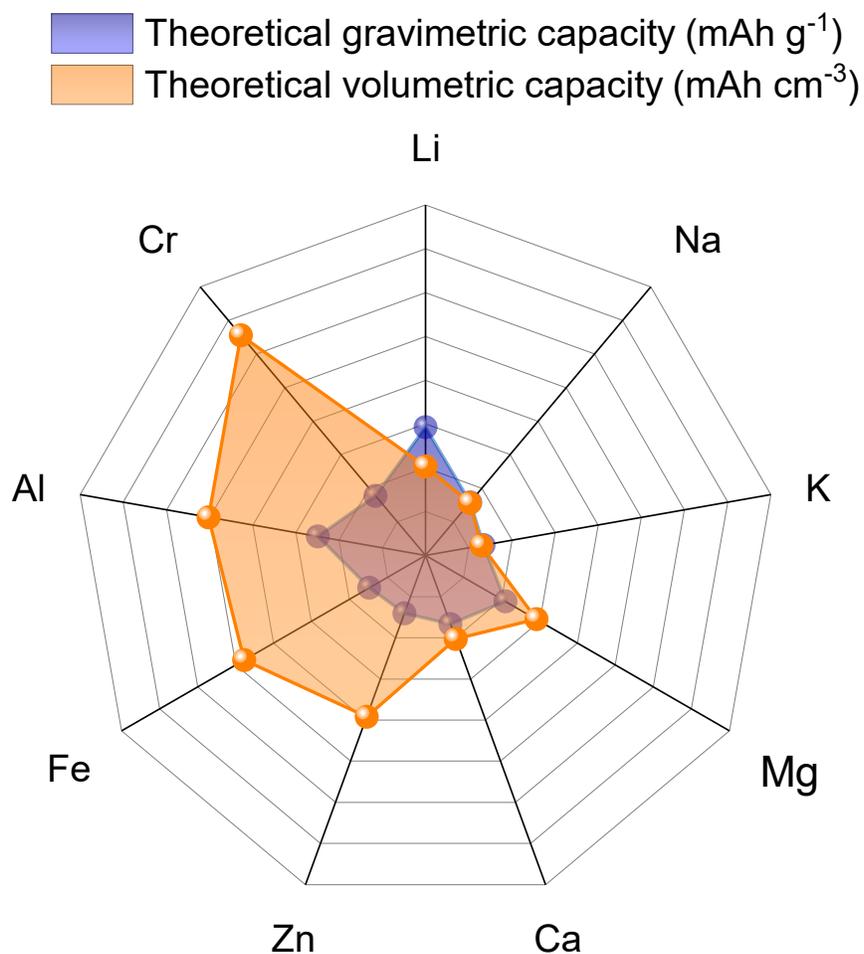

**Supplementary Figure S1 | Comparison of theoretical performance for metallic battery anodes.** A radar plot comparing the theoretical gravimetric capacity (mAh g$^{-1}$, blue) and theoretical volumetric capacity (mAh cm$^{-3}$, orange) for selected metallic elements as potential anodes in rechargeable batteries. Among the elements shown, chromium (Cr) possesses the highest theoretical volumetric capacity, calculated to be ~11,117 mAh cm$^{-3}$.

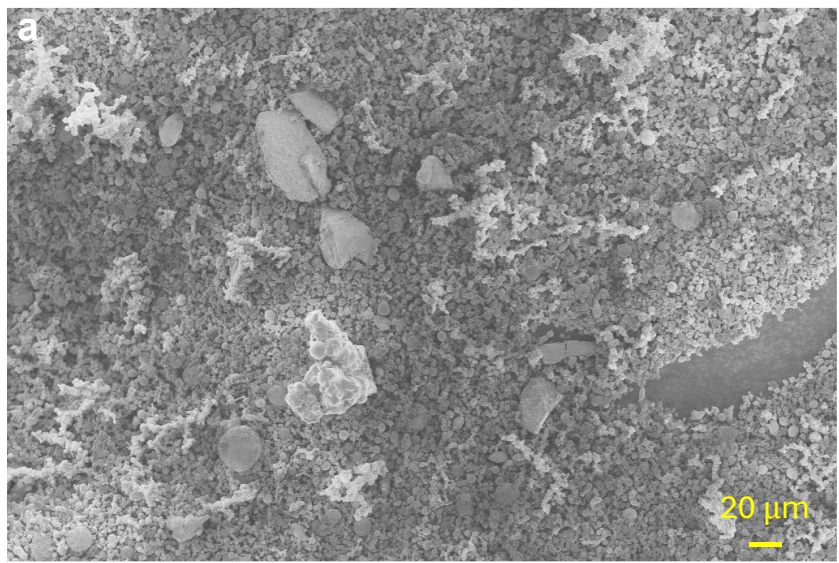
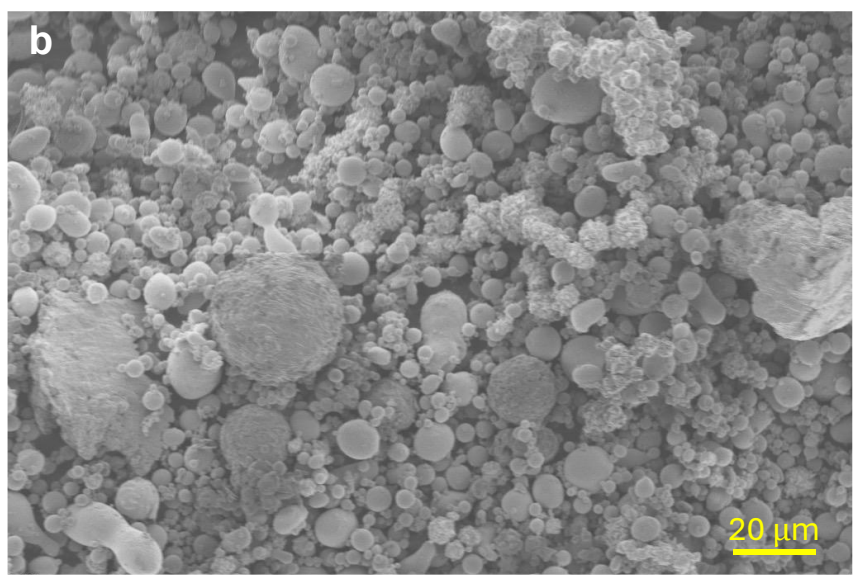

**Supplementary Figure S2 | Scanning electron microscopy images of elemental powder mixture before ball milling. a, b,** Low-magnification scanning electron microscopy (SEM) images of the mixed Cr, Bi, Cu, Ni, and Sn elemental powders. The powders were manually mixed with a spatula before being transferred for ball milling (0 h milling time). The images show a heterogeneous distribution of various particle sizes and morphologies.

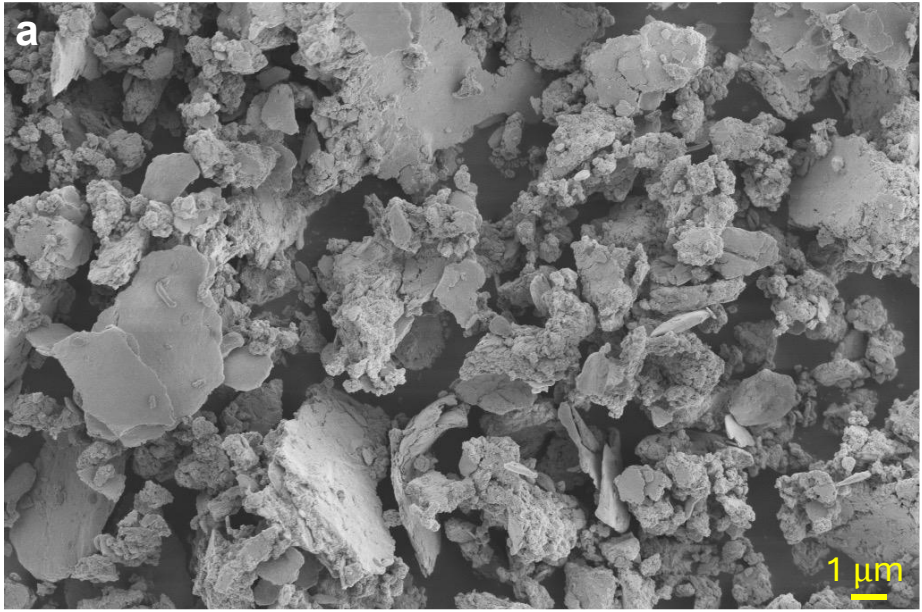
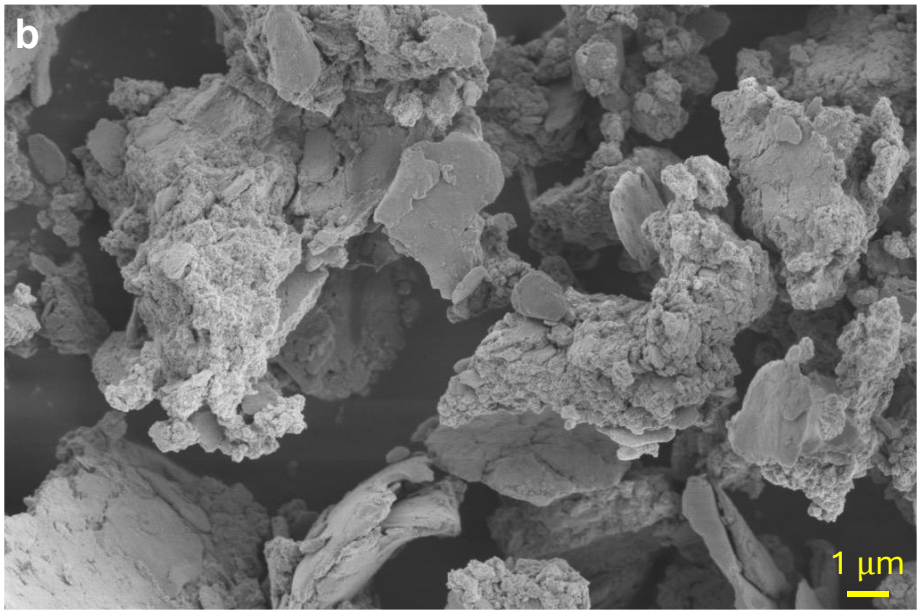

**Supplementary Figure S3 | Scanning electron microscopy images of Cr-based high-entropy alloy powder after ball milling. a, b,** Representative scanning electron microscopy (SEM) images of the Cr-based high-entropy alloy (Cr-HEA) powder. The powder was obtained after ~20 hours of continuous ball milling in toluene, followed by drying. The images reveal particles with an average size of ~1.9 μm.

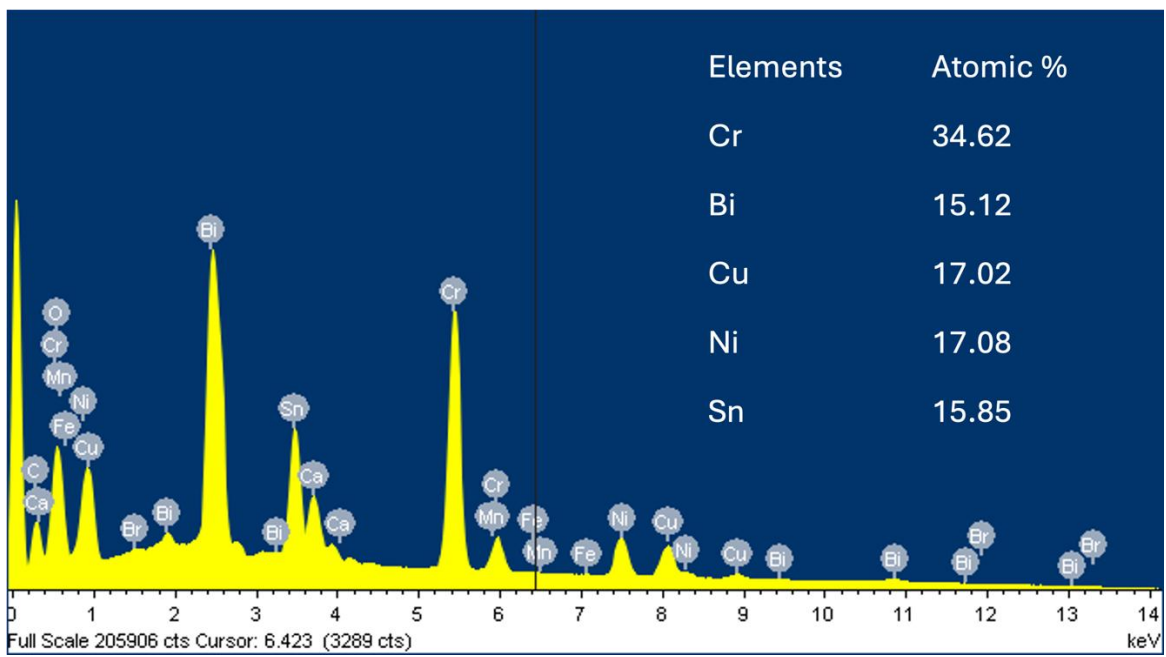

**Supplementary Figure S4 | Elemental analysis of the Cr-based high-entropy alloy powder.** Energy-dispersive X-ray spectroscopy (EDS) spectrum of the synthesized Cr-HEA powder. The inset table lists the quantified elemental composition in atomic percent (at. %). The analysis confirms the powder is composed of Cr (~34.62%), Cu (~17.02%), Ni (~17.08%), Sn (~15.85%), and Bi (~15.12%), validating the successful formation of the multi-component alloy.

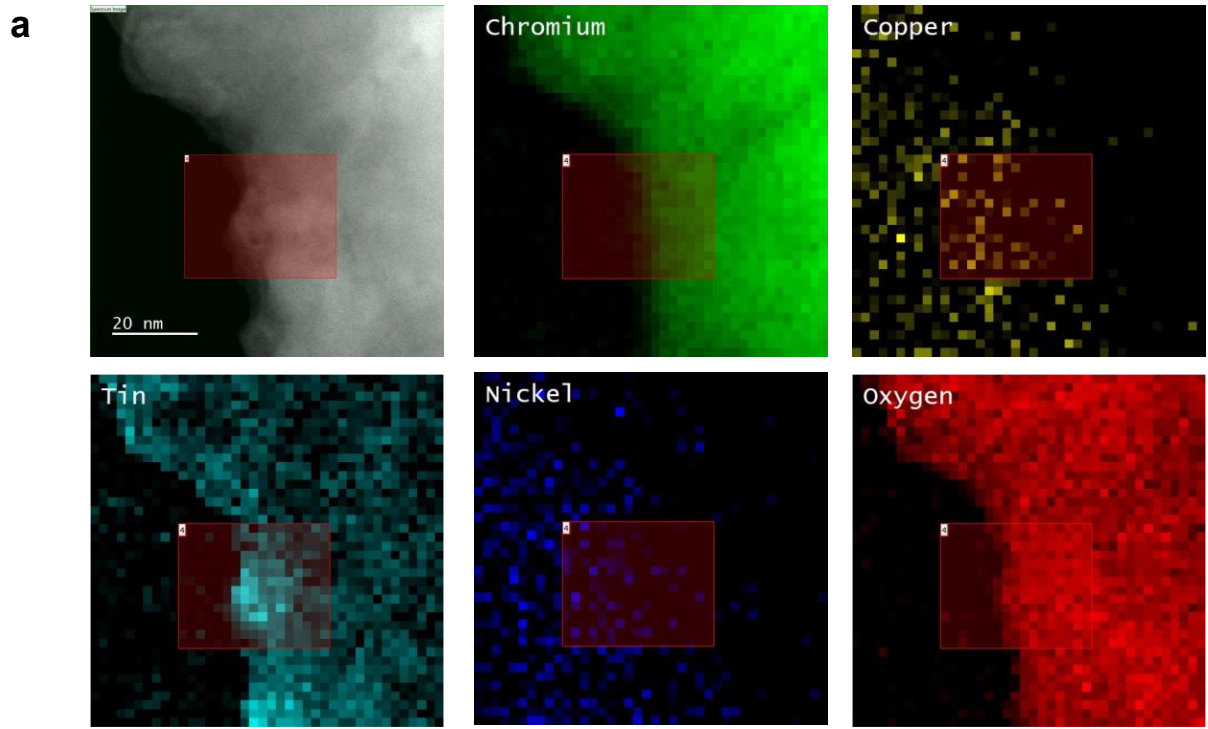

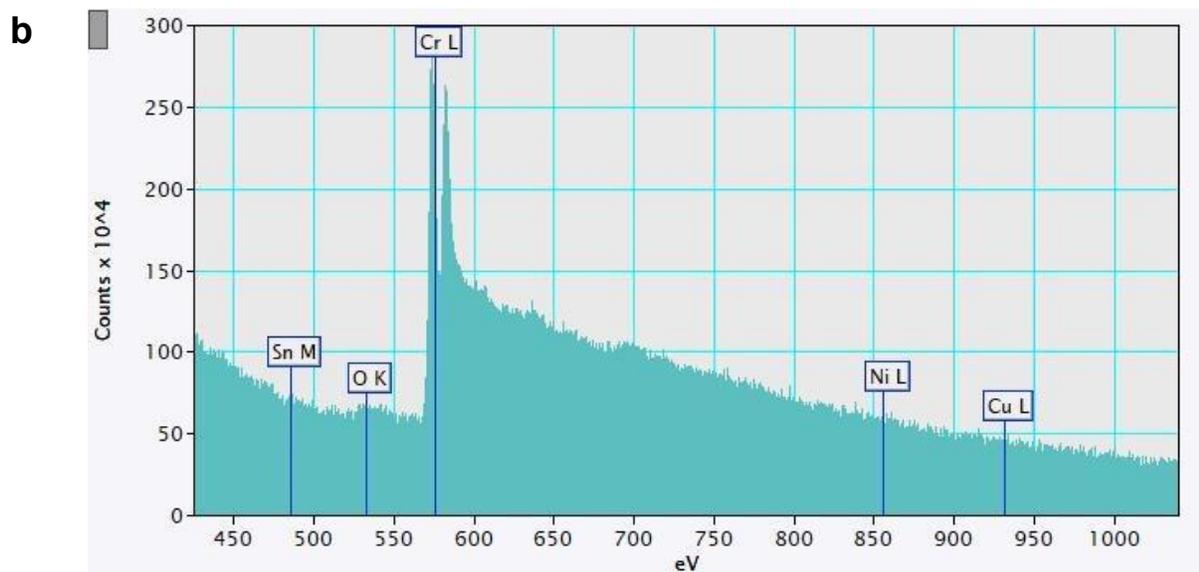

**Supplementary Figure S5 | Nanoscale compositional analysis of the Cr-HEA powder. a,** High-angle annular dark-field scanning transmission electron microscopy (HAADF-STEM) image and corresponding electron energy loss spectroscopy (EELS) elemental maps showing the distribution of Cr, Cu, Sn, Ni, and O. **b,** EELS spectrum acquired from the region highlighted by the red box in (a), confirming the presence of the constituent elements. The combined analysis reveals an oxygen-rich shell on the surface of the metallic alloy core. The scale bar in (a) is 20 nm.

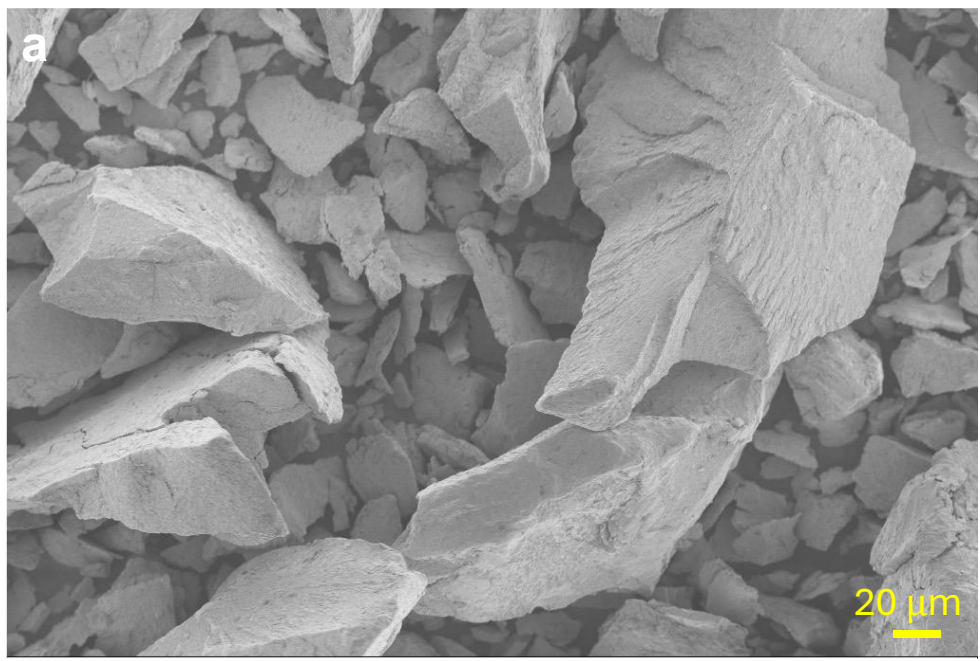

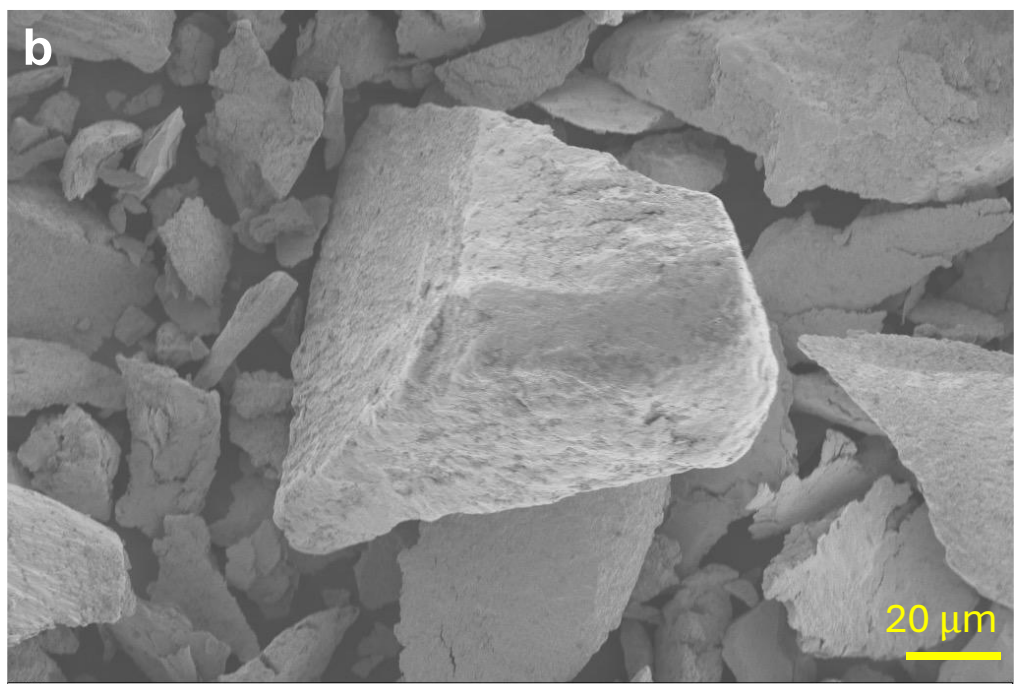

**Supplementary Figure S6 | Morphology of the as-received chromium powder. a, b,** Scanning electron microscopy (SEM) images of the as-received chromium (Cr) powder used for electrode fabrication without any pre-treatment. The images show large, pristine particles characterized by irregular and angular morphologies. The images reveal an average particle size of ~27 μm.

**a**

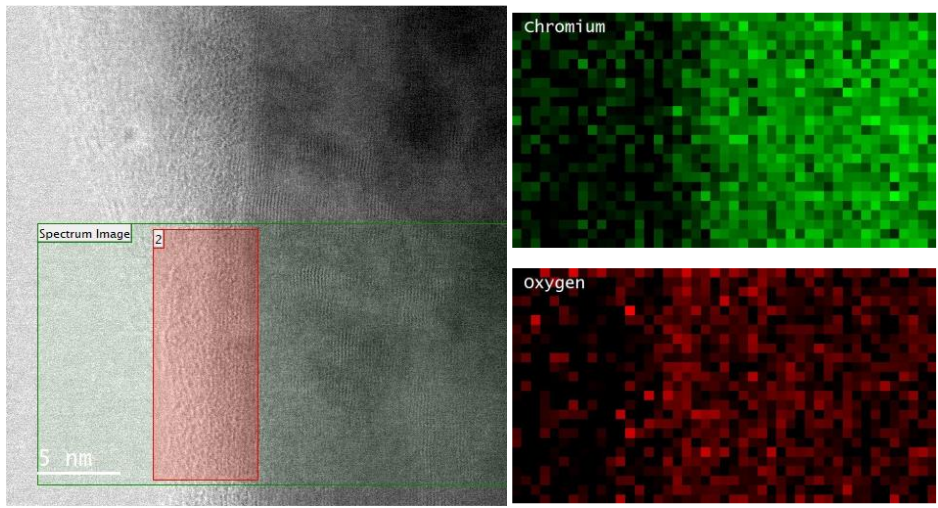

**b**

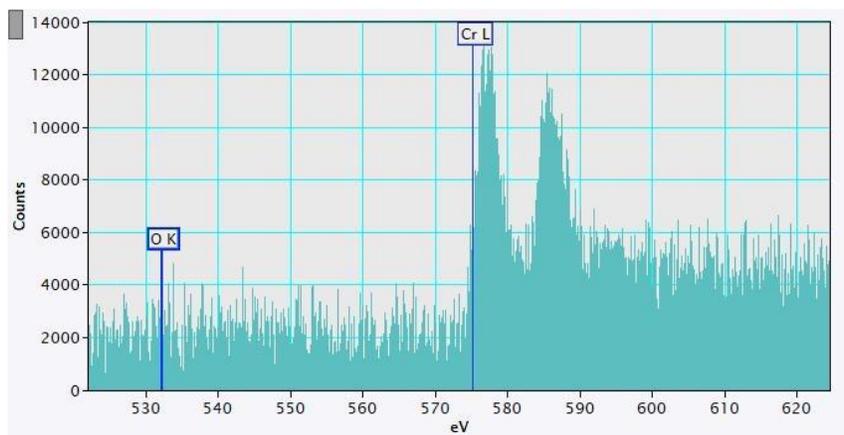

**Supplementary Figure S7 | EELS analysis of the Cr/Cr-oxide interface. a,** Cross-sectional high-angle annular dark-field scanning transmission electron microscopy (HAADF-STEM) image and corresponding EELS elemental maps for Chromium (green) and Oxygen (red). **b,** The EELS spectrum acquired from the interface region, indicated by the red box in (a). The analysis confirms the formation of a distinct chromium oxide layer adjacent to the metallic chromium. The scale bar in (a) is 5 nm.

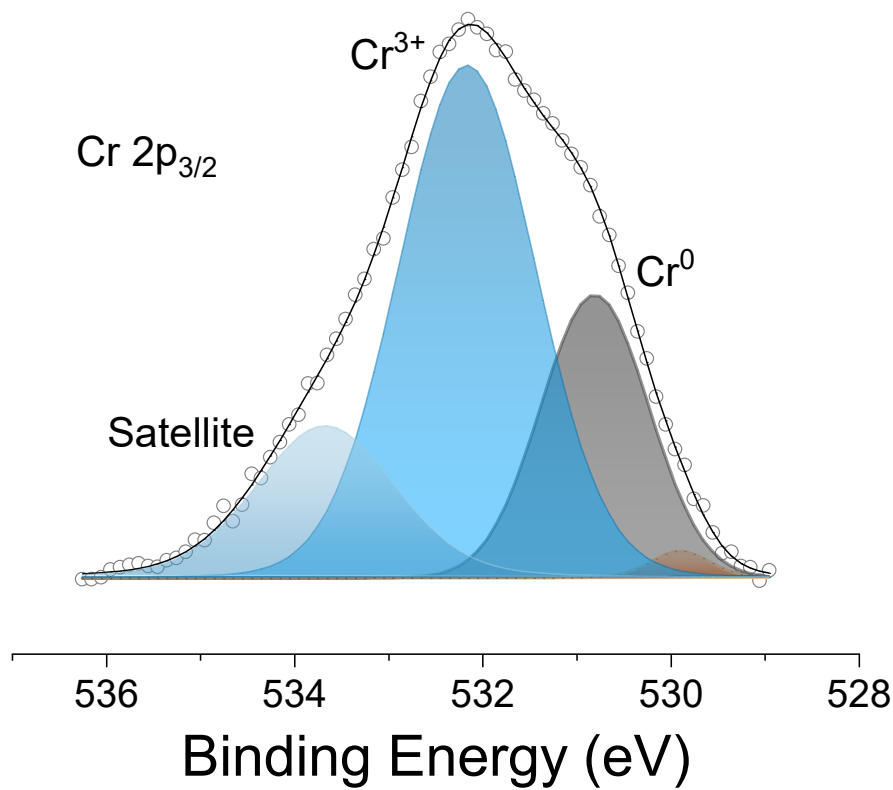

**Supplementary Figure S8 | Surface chemical state analysis of the fresh Cr electrode.** High-resolution X-ray photoelectron spectroscopy (XPS) spectrum of the Cr $2p_{3/2}$ core level for the fresh Cr electrode. The spectrum is deconvoluted into components corresponding to metallic chromium ($Cr^0$), chromium(III) oxide ($Cr_2O_3$), and a satellite peak that corresponds to chromium hydroxide. The presence and dominance of the oxide states confirm a native passive oxide layer on the surface of the pristine electrode.

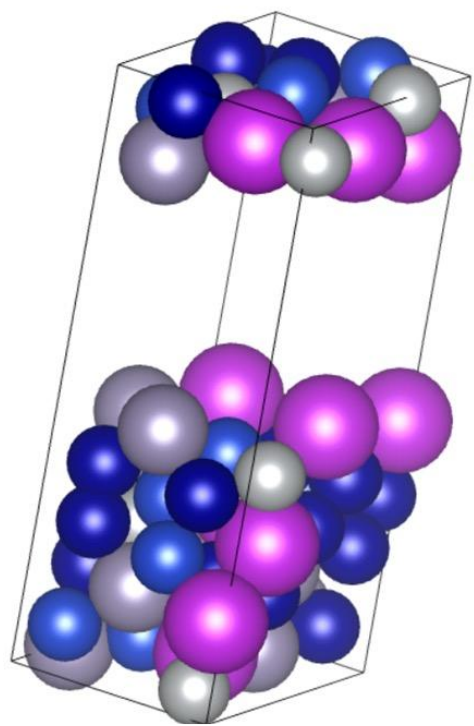

**Supplementary Figure S9 | Optimized structure of Cr-HEA (110) surface.**

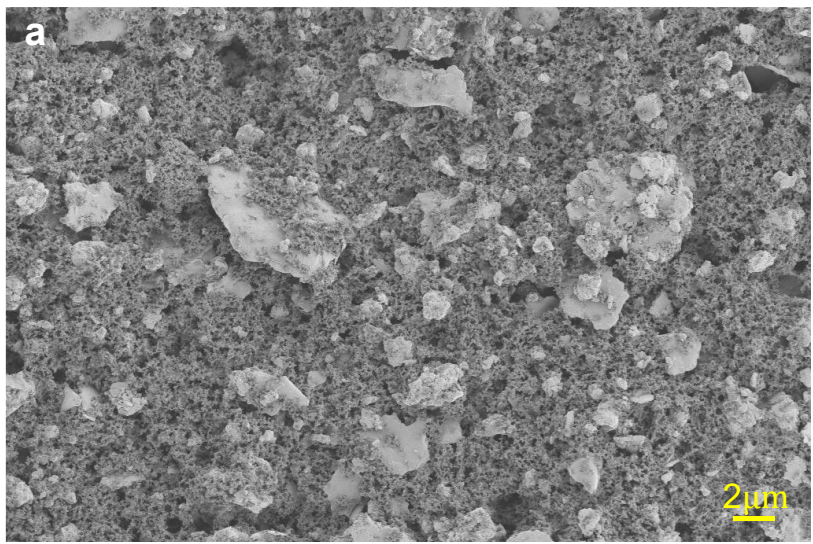
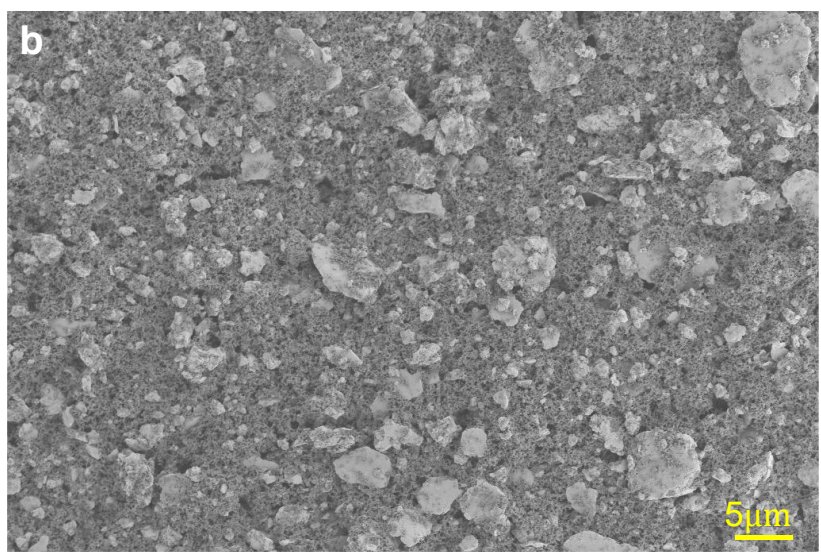

**Supplementary Figure S10 | Microstructure of the as-prepared Cr-HEA electrode. a, b,** Scanning electron microscopy (SEM) images of the Cr-HEA electrode surface at different magnifications. The micrographs show the Cr-HEA active material particles embedded within a porous, web-like structure. This interconnected network is formed by the polyvinylidene fluoride (PVDF) binder, which maintains the structural integrity of the electrode.

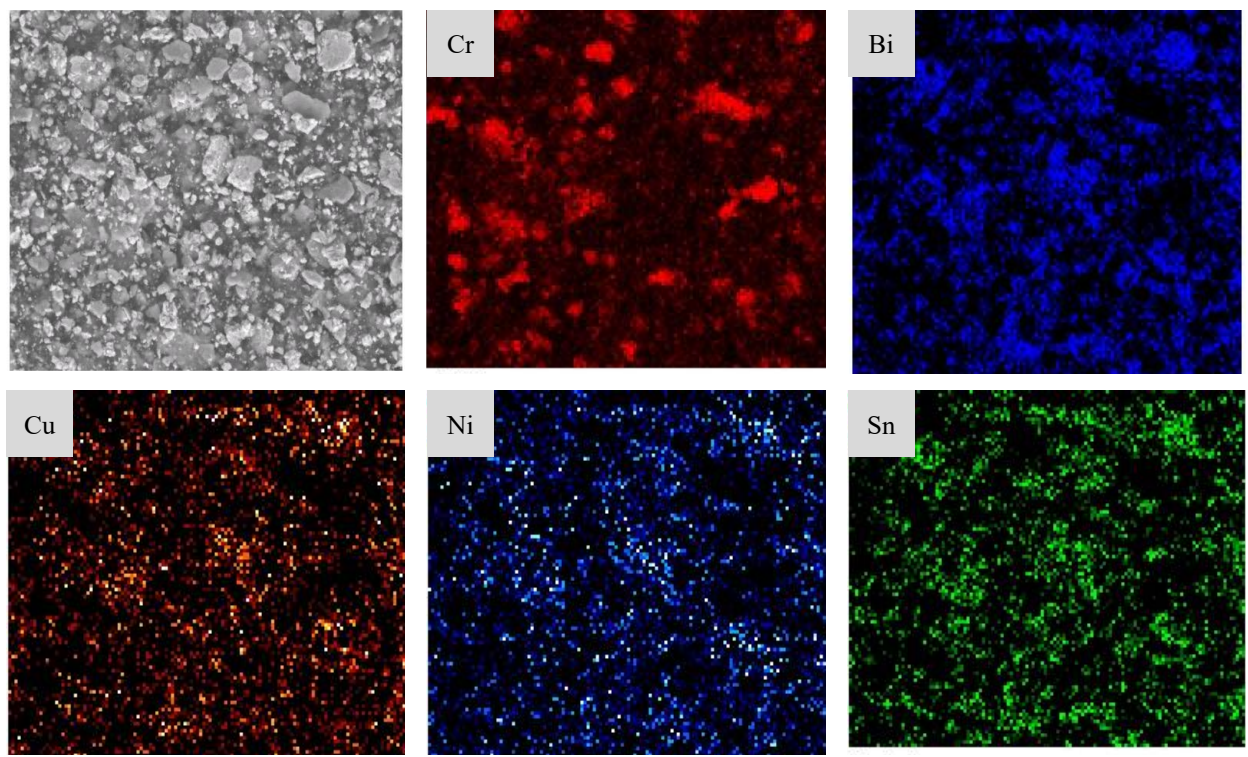

**Supplementary Figure S11 | Elemental distribution in the Cr-HEA electrode.** Scanning electron microscopy (SEM) image and corresponding energy-dispersive X-ray spectroscopy (EDS) elemental maps of the Cr-HEA electrode surface. The maps show the distribution of the primary constituent elements: Cr, Bi, Cu, Ni, and Sn. The uniform co-location of these elements confirms that the homogeneous composition of the high-entropy alloy powder is well-preserved throughout the electrode structure after fabrication.

a

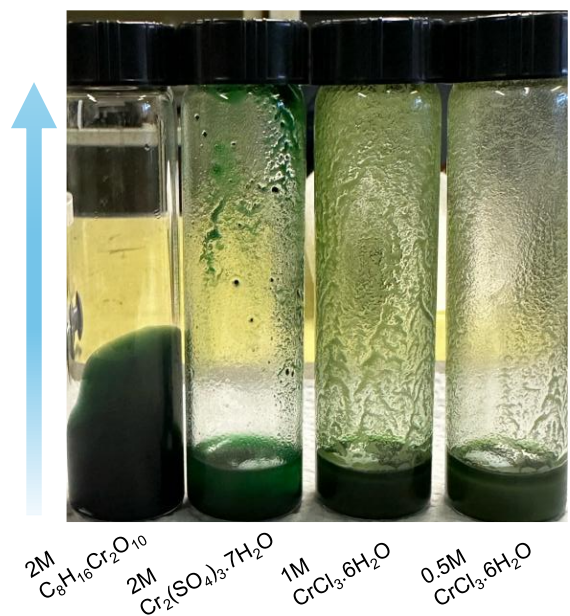

b

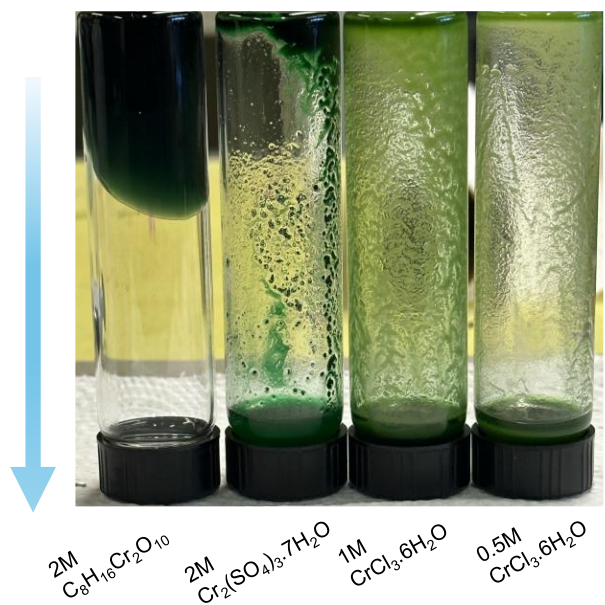

**Supplementary Figure S12 | Stability of various chromium salt electrolytes in DMSO.** Digital photographs of electrolytes prepared by dissolving different chromium salts in DMSO, pictured after resting for three days under an argon atmosphere. From left to right, the vials contain 2.0 M chromium(II) acetate, 2.0 M chromium(III) sulfate, 1.0 M chromium(III) chloride, and 0.5 M chromium(III) chloride. Panels **a** and **b** document the varying solubility of the salts, highlighting the significant precipitation in the acetate and sulfate electrolytes compared to the more stable chloride-based solutions.

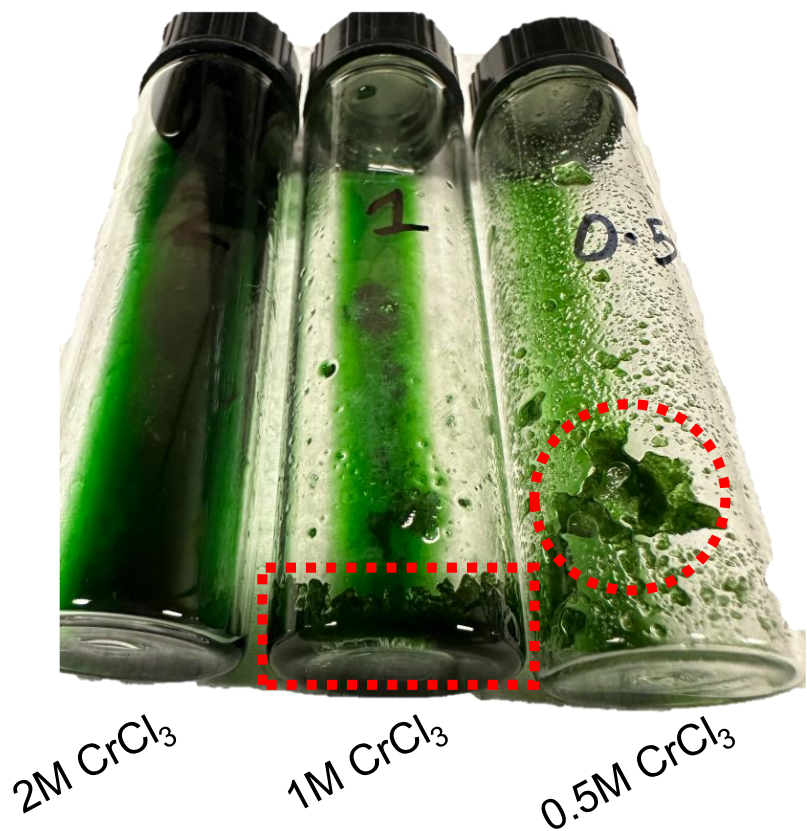

**Supplementary Figure S13 | Concentration-dependent stability of CrCl₃ in DMSO electrolyte.** Digital photographs showing the long-term stability of CrCl₃ dissolved in DMSO at various concentrations after three days under argon. From left to right: 2.0 M CrCl₃, 1.0 M CrCl₃, and 0.5 M CrCl₃. The red dashed boxes highlight the significant formation of insoluble byproducts in the 1.0 M and 0.5 M solutions, indicating reduced stability at lower concentrations. In contrast, the 2.0 M CrCl₃ electrolyte remains visually clear, suggesting superior stability at higher concentrations.

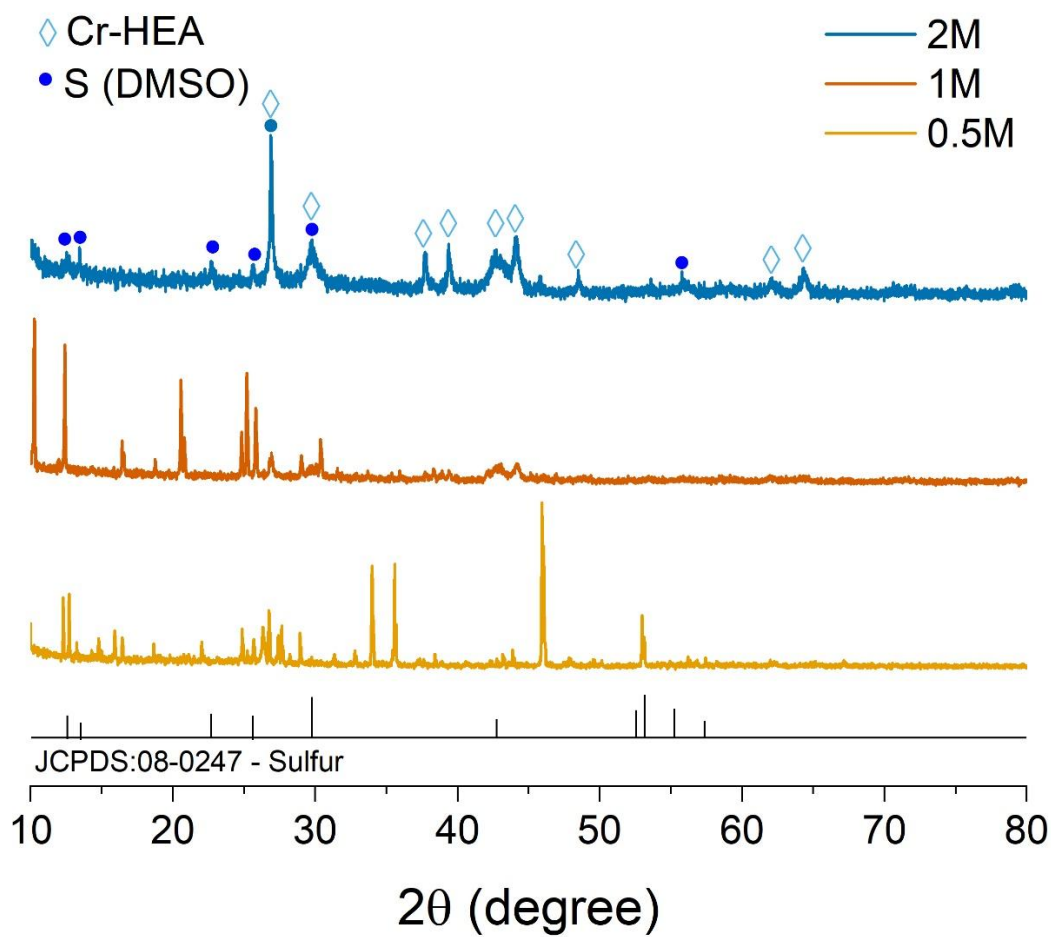

**Supplementary Figure S14 | Stability analysis of the Cr-HEA electrode in different electrolyte concentrations.** X-ray diffraction (XRD) patterns of Cr-HEA electrodes after immersion in $CrCl_3·6H_2O$/DMSO electrolytes with concentrations of 2.0 M, 1.0 M, and 0.5 M for five days. The patterns for the electrodes aged in the 1.0 M and 0.5 M electrolytes exhibit numerous additional diffraction peaks, indicating the formation of crystalline byproducts from a reaction with the electrolyte. In contrast, the pattern for the electrode aged in the 2.0 M electrolyte primarily shows peaks corresponding to the original Cr-HEA phase, confirming its superior chemical stability in the higher concentration solution.

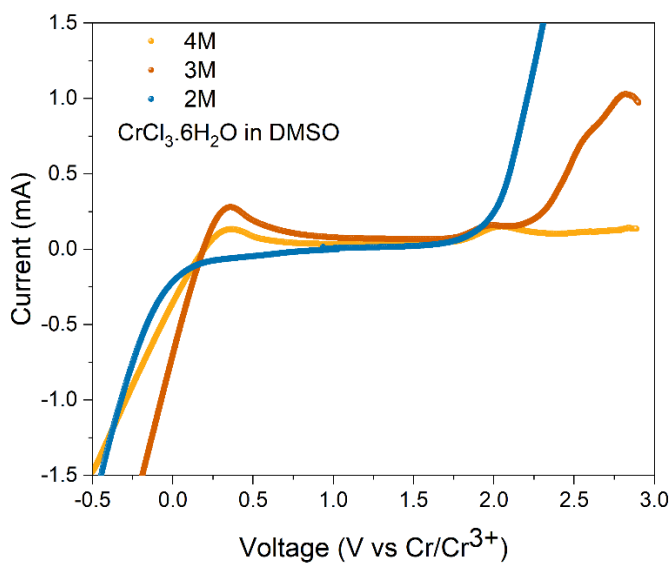

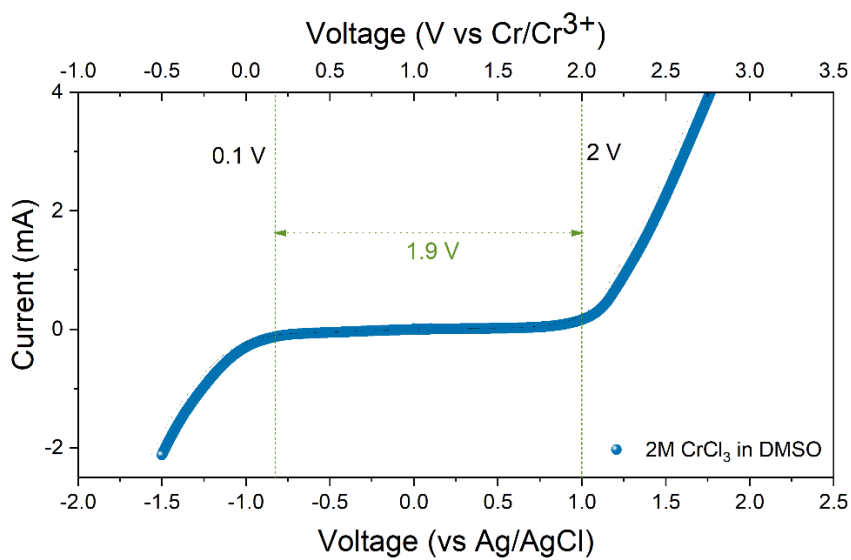

**Supplementary Figure S15 | Electrochemical stability window of the CrCl₃/DMSO electrolyte.** Linear sweep voltammetry (LSV) curves measured using a three-electrode setup with a titanium foil working electrode, a platinum counter electrode, and an Ag/AgCl reference electrode. **a,** Comparison of LSV curves for electrolytes containing 2.0 M, 3.0 M, and 4.0 M $CrCl_3·6H_2O$ in DMSO. The potential scale has been converted to V vs. $Cr/Cr^{3+}$. **b,** LSV curve for the 2.0 M electrolyte, which demonstrates an electrochemical stability window of ~1.9 V.

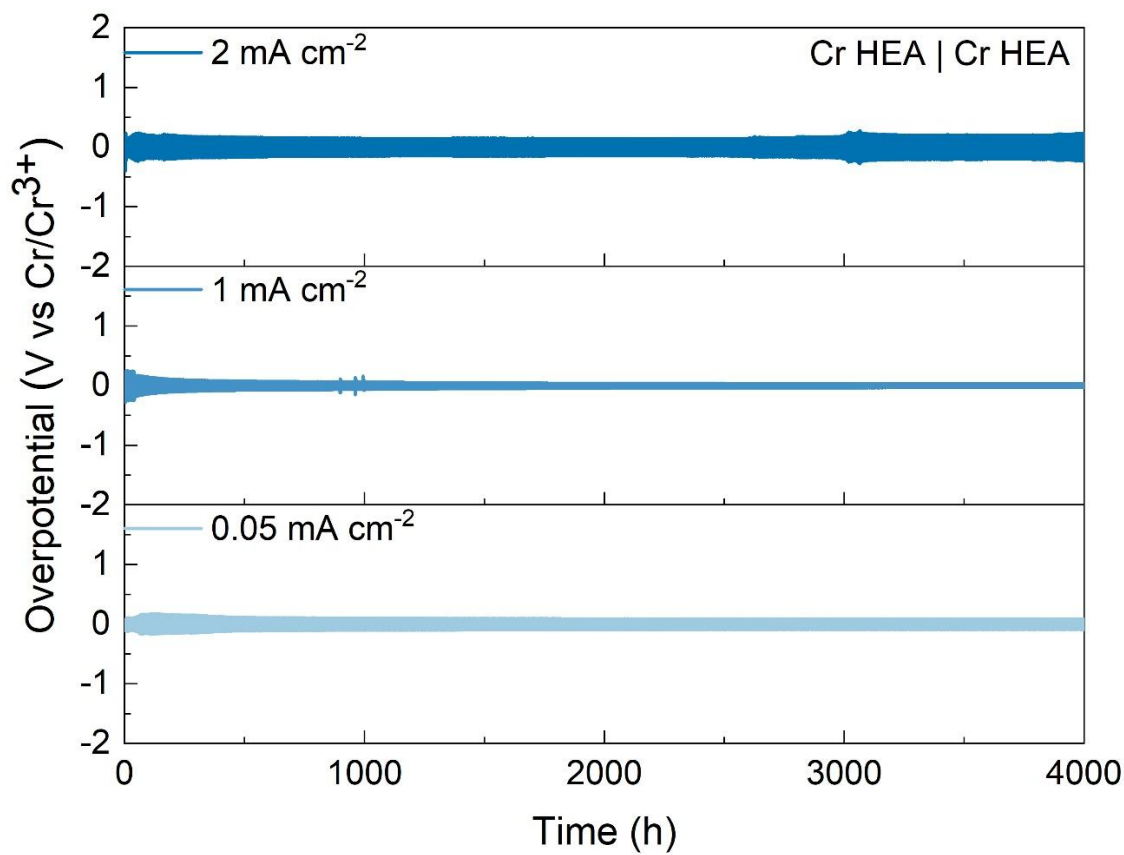

**Supplementary Figure S16 | Long-term $Cr^{3+}$ shuttling stability of Cr-HEA symmetric cells.** Time-dependent overpotential profiles of symmetric cells with Cr-HEA electrodes cycled at current densities of ~0.05, ~1.0, and ~2.0 mA cm$^{-2}$. The cells exhibit stable response for over 4000 h with minimal polarization, demonstrating the ability of the Cr-HEA anode to sustain prolonged Cr-ion transport across a wide range of rates.

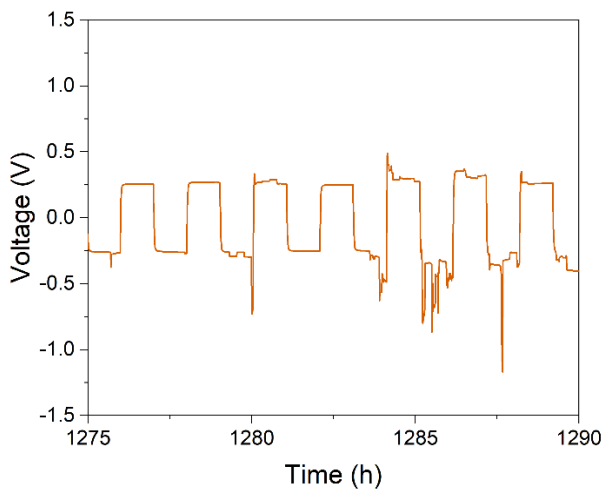

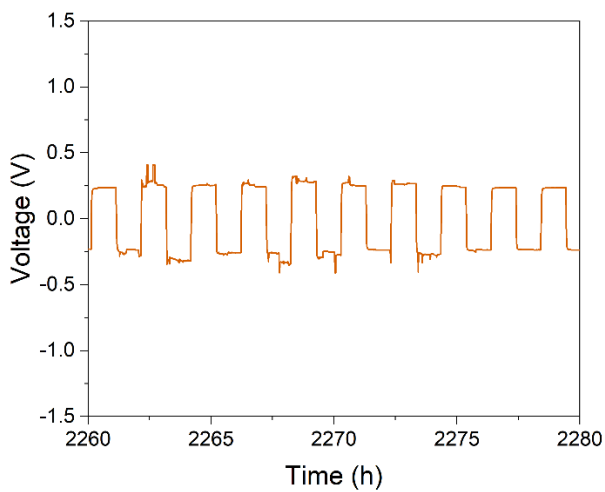

**Supplementary Figure S17 | Evolution of Cr plating and stripping stability in a symmetric cell.** Voltage profiles of a Cr||Cr symmetric cell during galvanostatic cycling at different operational stages. **a,** an early-stage profile (~1280 h) exhibiting unstable and asymmetric behavior with significant voltage fluctuations, characteristic of inefficient plating and stripping processes. **b,** A later-stage profile (~2265 h) showing a similar behavior.

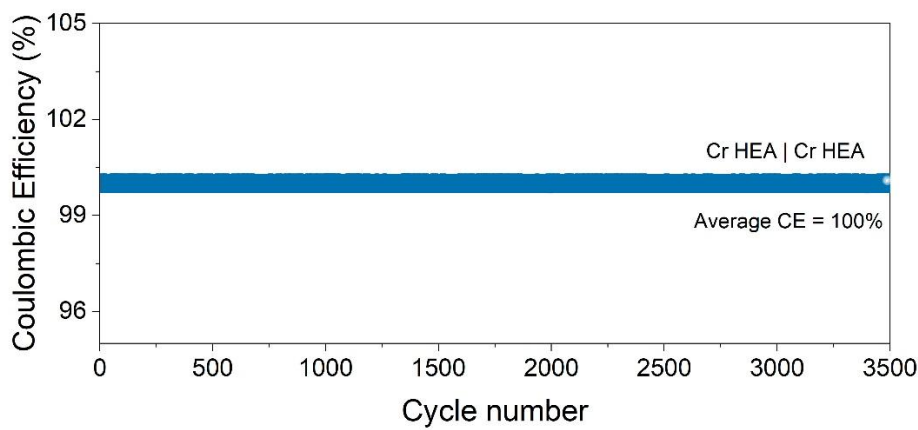

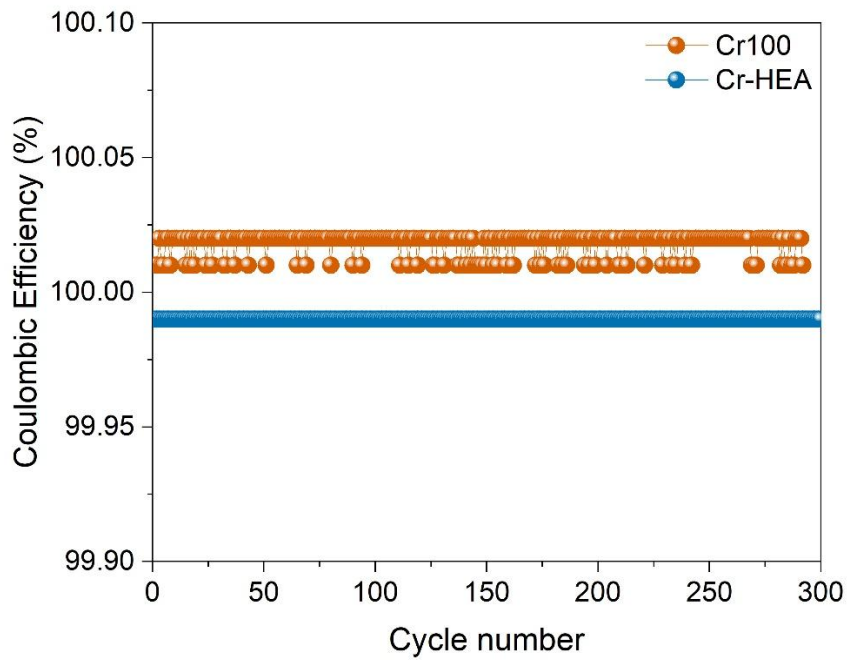

**Supplementary Figure S18 | Coulombic efficiency (CE) for $Cr^{3+}$ shuttling measured at a current density of ~0.1 mA cm$^{-2}$. a**, Long-term cycling performance using a Cr-HEA symmetric cell, which demonstrates a highly stable CE averaging (nearly 100%) over 3500 cycles (approximately 7000 hours). **b,** A magnified view of the initial 100 cycles comparing the stability of a Cr-HEA || Cr-HEA cell (orange) with a Cr || Cr cell (pink). The Cr-HEA electrode enables exceptionally stable and consistent CE from the onset, whereas the bare Cr electrode results in significant fluctuations during the initial cycles.

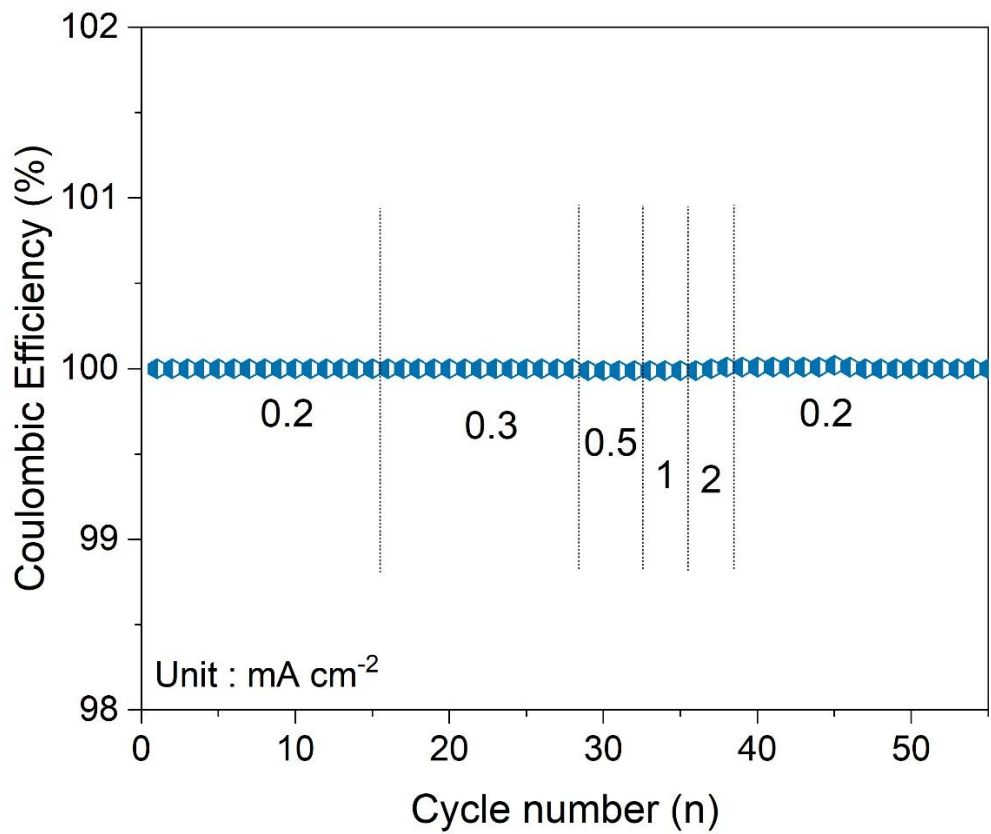

**Supplementary Figure S19 | Rate performance of the Cr-HEA electrode.** Coulombic efficiency (CE) of the Cr-HEA electrode at various current densities for a fixed areal capacity of ~1 mAh cm$^{-2}$. The current density was incrementally increased from ~0.2 to ~2.0 mA cm$^{-2}$ and subsequently returned to ~0.2 mA cm$^{-2}$. The electrode maintains a stable CE of approximately 100% across all tested rates, highlighting its excellent reversibility of Cr$^{3+}$ shuttling.

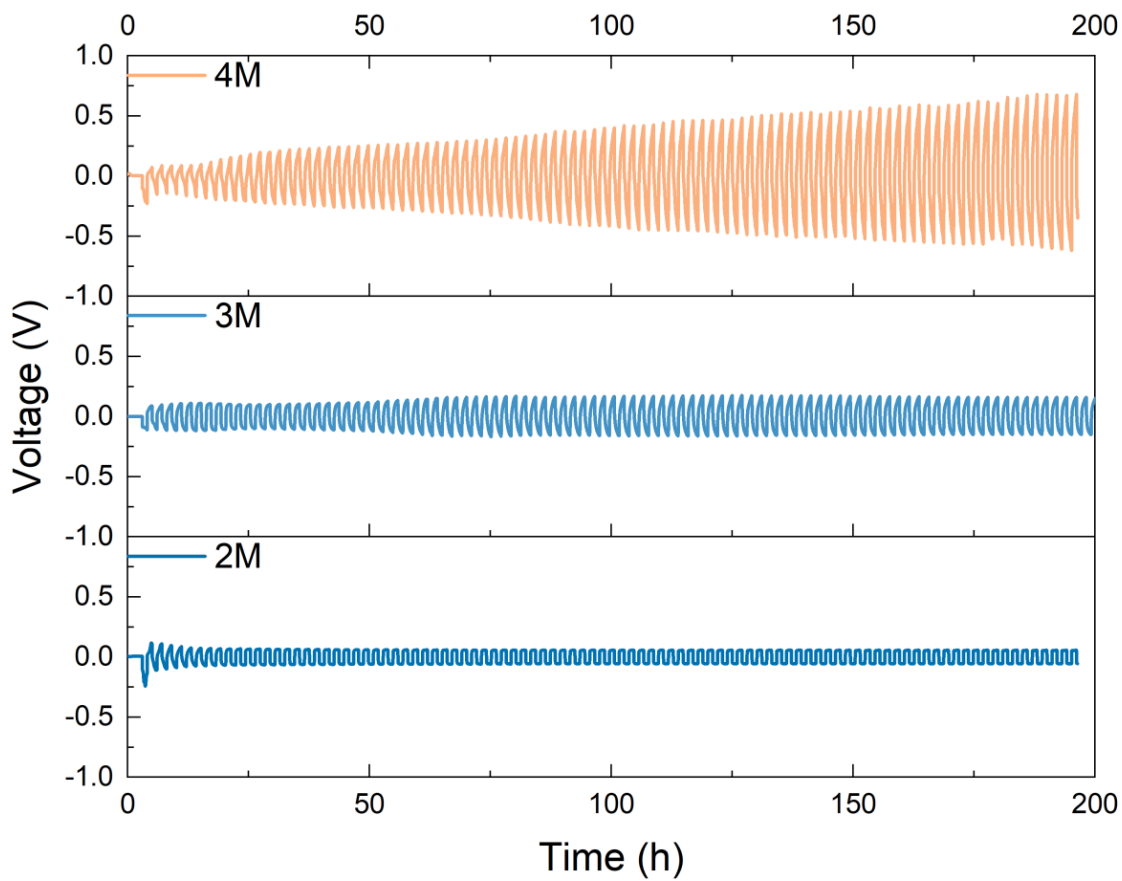

**Supplementary Figure S20 | Effect of electrolyte concentration on $Cr^{3+}$ shuttling stability.** Voltage profiles from the galvanostatic cycling of Cr-HEA || Cr-HEA symmetric cells using $CrCl_3$/DMSO electrolytes at concentrations of ~4.0 M, ~3.0 M, and ~2.0 M. The ~2.0 M electrolyte exhibits the lowest and most stable overpotential over 200 hours of cycling. In contrast, higher concentrations lead to increased overpotentials and instability, particularly for the ~4.0 M solution. Based on this superior performance, the ~2.0 M concentration was selected as the optimal electrolyte for this study.

a

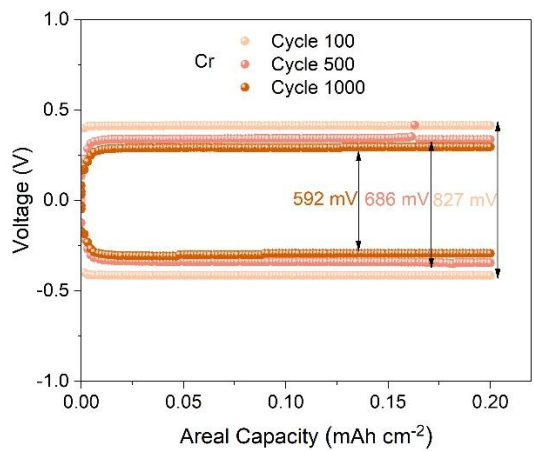

b

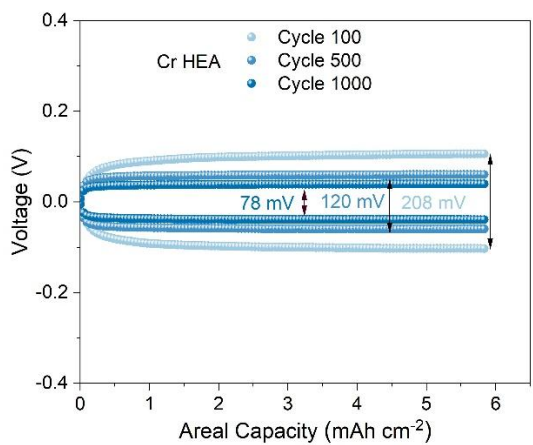

**Supplementary Figure S21 | Performance comparison in Cr vs. Cr-HEA symmetric cells**. Voltage vs. areal capacity profiles at the 100th, 500th, and 1000th cycles. **a**, A standard Cr||Cr symmetric cell, which exhibits a large and increasing voltage hysteresis (overpotential) while being limited to a low areal capacity of ~0.2 mAh cm$^{-2}$, indicating severe interfacial resistance. **b**, A Cr-HEA||Cr-HEA symmetric cell, demonstrating significantly smaller and more stable voltage hysteresis. This configuration achieves a high areal capacity of ~6.0 mAh cm$^{-2}$, confirming greatly improved interfacial kinetics and efficient Cr$^{3+}$ shuttling.

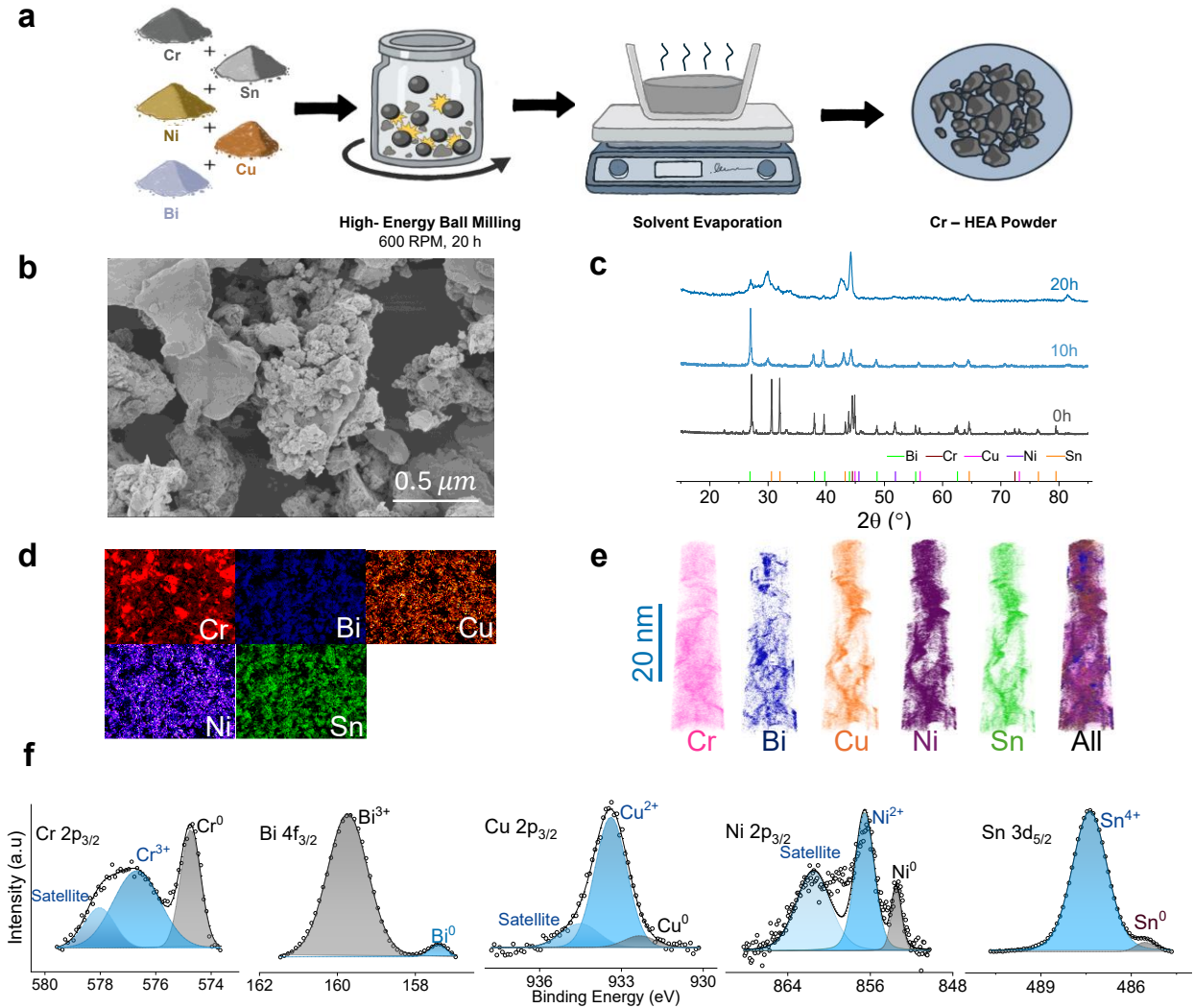

**Fig. 1 | Synthesis and structural characterization of the Cr-based high-entropy alloy. a,** Schematic illustration of the top-down synthesis of the $Cr_{0.4}Bi_{0.15}Cu_{0.15}Ni_{0.15}Sn_{0.15}$ high-entropy alloy (Cr-HEA) via high-energy ball milling. **b,** Scanning electron microscopy (SEM) image of the as-synthesized Cr-HEA powder, revealing an average particle size of ~1.9 ± 0.5 μm. **c,** Evolution of X-ray diffraction (XRD) patterns over various milling times. The disappearance of distinct elemental peaks and the emergence of a dominant diffraction peak after 20 h indicate the successful formation of a single-phase solid solution. **d,** Elemental dispersive spectroscopy (EDS) analysis of the as-synthesized Cr HEA powder. **e,** Atom probe tomography (APT) reconstruction providing a three-dimensional elemental map of a representative nanoscale volume. The analysis confirms the incorporation of all five elements (Cr, Bi, Cu, Ni, Sn) into the solid solution, while highlighting the local atomic-scale heterogeneities characteristic of ball-milled alloys. **f,** High-resolution X-ray photoelectron spectroscopy (XPS) spectra of the constituent elements. The analysis reveals the presence of a native, mixed-cation oxide layer on the alloy surface, distinct from the simple oxide layer found on pure chromium.

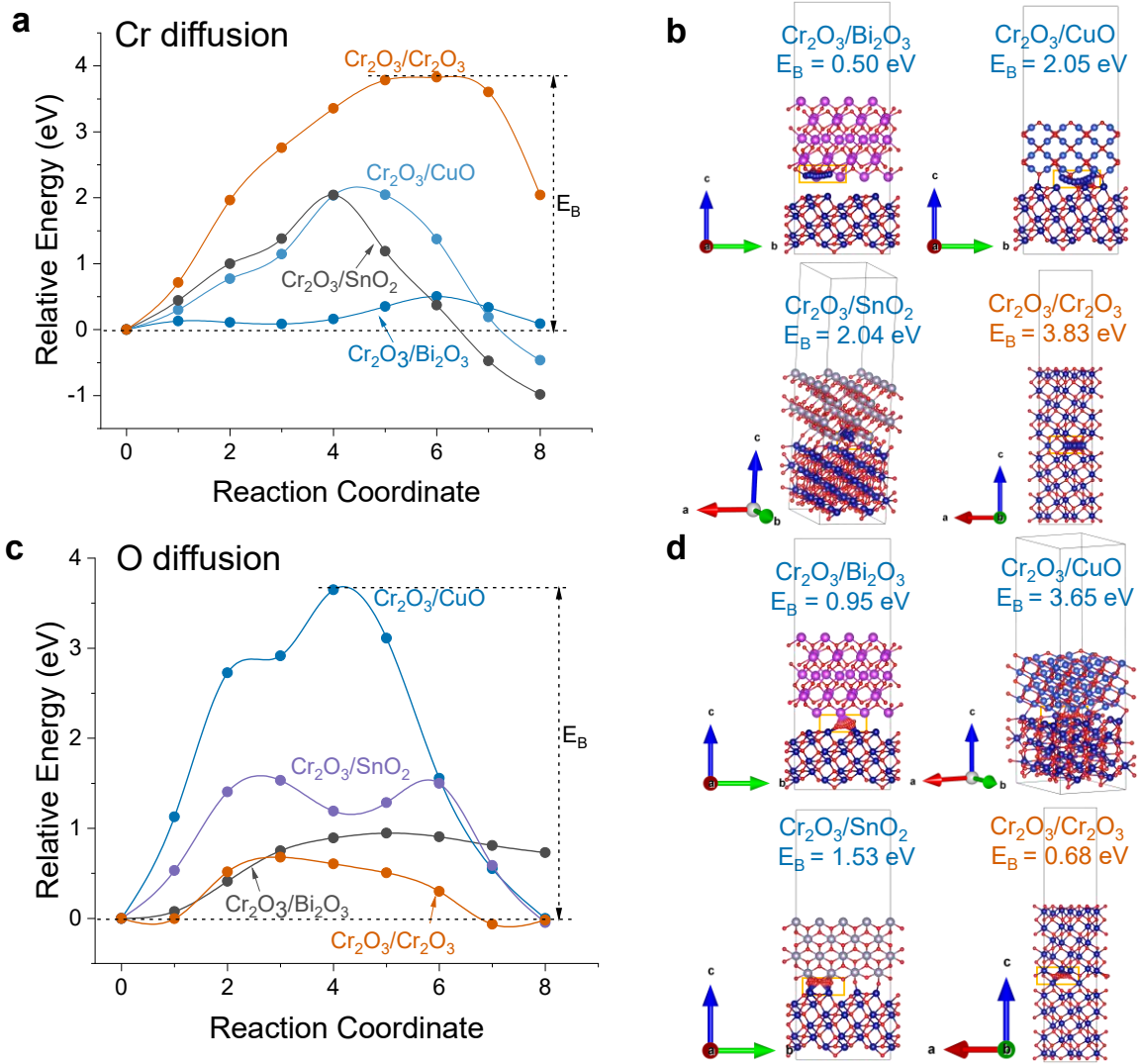

**Fig.2 | Computational study of Cr and O diffusion through the multi-element oxide interface in Cr-HEA**. **a,** NEB results for migration of Cr along $Cr_2O_3/Bi_2O_3$, $Cr_2O_3/SnO_2$, $Cr_2O_3/CuO$ hetero-interfaces and $Cr_2O_3/Cr_2O_3$ grain boundaries. **b,** Migration path of Cr for the interfaces in a, with the corresponding migration energy barrier ($E_B$). **c,** NEB results for migration of O along $Cr_2O_3/Bi_2O_3$, $Cr_2O_3/SnO_2$, $Cr_2O_3/CuO$ hetero-interfaces and $Cr_2O_3/Cr_2O_3$ grain boundaries. **d,** Migration path of O for the interfaces in c, with the corresponding migration energy barrier ($E_B$).

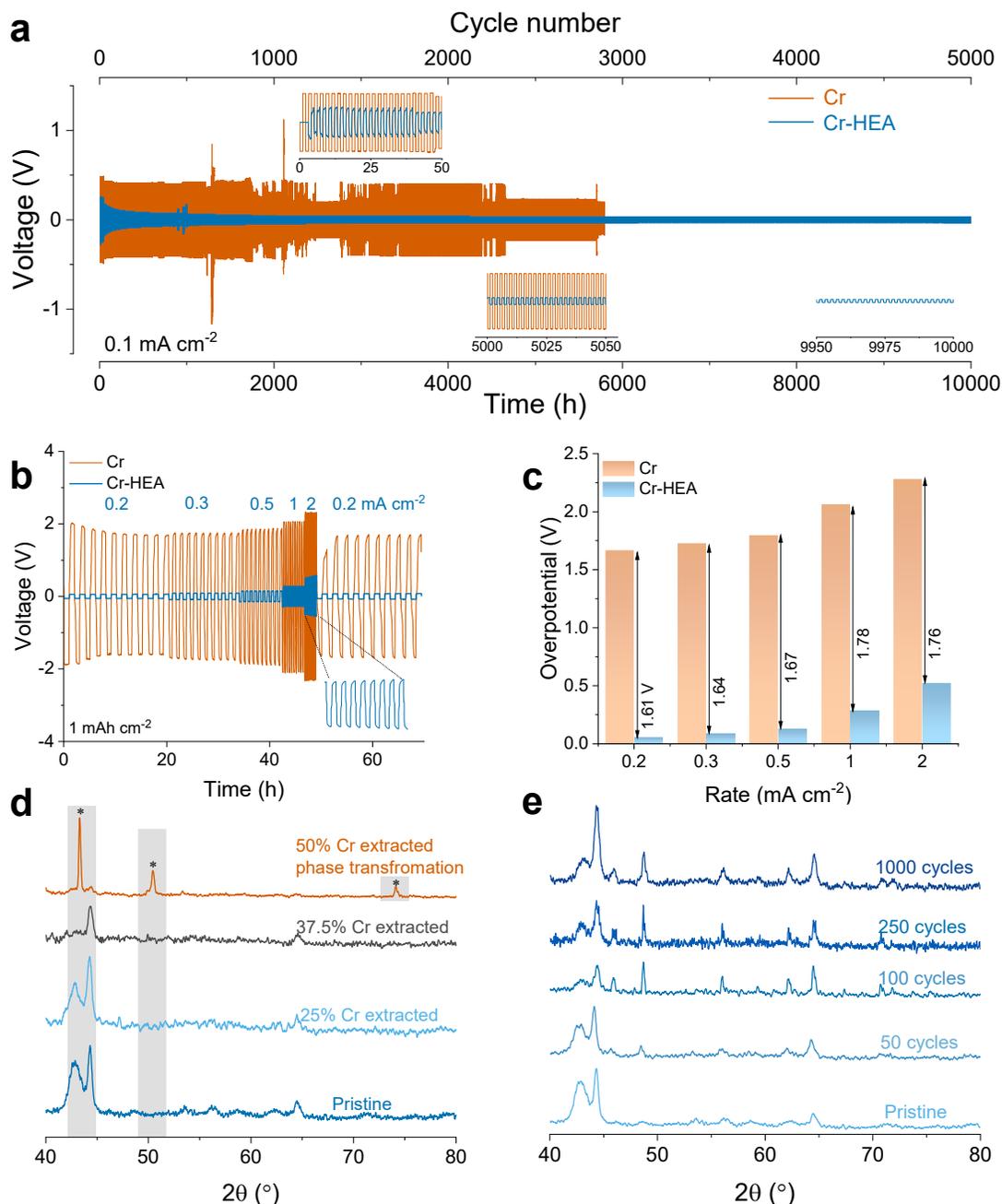

**Fig. 3 | Comparison of Cr–HEA and pure Cr metal electrodes in symmetric cells. a,** Long-term $Cr^{3+}$ insertion/extraction in symmetric cells using Cr–HEA (blue) and pure Cr (orange) electrodes. **b,** Response of Cr–HEA and Cr symmetric cells under varying current densities. **c,** Overpotential differences between Cr-HEA and Cr symmetric cells at different current densities. **d,** One-way transfer experiments indicating that up to ~37.5% of Cr can be extracted without significantly disrupting the Cr-HEA structure, whereas ~50% extraction induces a new phase. **e,** XRD patterns confirm that the Cr-HEA framework remains intact even after 1000 continuous $Cr^{3+}$ insertion/extraction cycles, underscoring its structural stability.

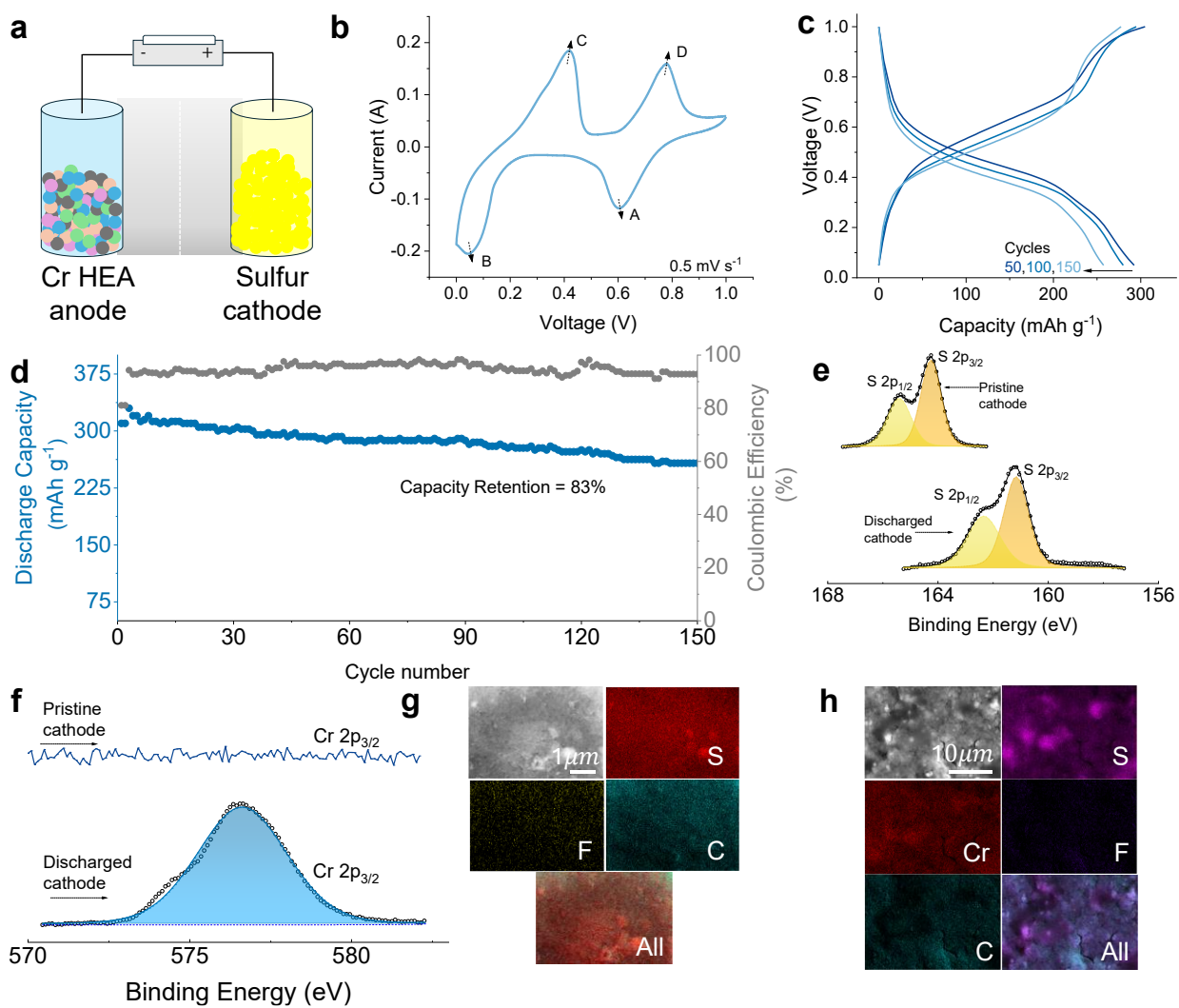

**Fig. 4 | Electrochemical performance of Cr-HEA || S full-cell battery. a,** Schematic illustration of the full-cell configuration comprising the Cr-HEA anode and a sulfur/carbon composite cathode. **b,** Cyclic voltammetry (CV) profile of the full cell at a scan rate of ~0.5 mV s$^{-1}$ in the voltage range of 0.05–1.0 V. The redox peaks are labeled A–D. **c,** Galvanostatic charge/discharge voltage profiles for the 50$^{th}$, 100$^{th}$, and 150$^{th}$ cycles at a current density of ~100 mA g$^{-1}$. **d,** Cycling performance showing discharge capacity and corresponding coulombic efficiency at ~100 mA g$^{-1}$. **e, f,** High-resolution *ex situ* XPS spectra of the sulfur cathode in pristine and fully discharged states for the **(e)** S 2p region and **(f)** Cr 2p region. The emergence of Cr signals and the shift in S peaks confirm the Cr-S conversion reaction. **g, h,** SEM images and corresponding EDS elemental mapping for the **(g)** pristine sulfur cathode and **(h)** fully discharged sulfur cathode, showing the uniform distribution of migrated chromium after discharge (F in the maps originates from the PVDF binder).